\documentclass{emulateapj}
\slugcomment{Submitted to \textsc{the Astrophysical Journal}}
\shorttitle{Sedimentation and X-ray Bursts}
\shortauthors{Peng et al.}
\newcommand{\unitspace}{\ensuremath{\,}}
\newcommand{\usp}{\unitspace}
\newcommand{\numberspace}{\ensuremath{\;}}
\newcommand{\nsp}{\numberspace}
\newcommand{\unitstyle}[1]{\ensuremath{\mathrm{#1}}}
\newcommand{\power}[2]{\ensuremath{{#1}^{#2}}}

\newcommand{\kilo}{\unitstyle{k}}
\newcommand{\Mega}{\unitstyle{M}}

\newcommand{\cm}{\unitstyle{cm}}
\newcommand{\gram}{\unitstyle{g}}

\newcommand{\meter}{\unitstyle{m}}

\newcommand{\second}{\unitstyle{s}}

\newcommand{\Kelvin}{\unitstyle{K}}
\newcommand{\K}{\Kelvin}  

\newcommand{\grampercc}{\gram\usp\power{\cm}{-3}} 
\newcommand{\grampersquarecm}{\gram\usp\power{\cm}{-2}} 

\newcommand{\GramPerSc}{\grampersquarecm}
\newcommand{\erg}{\unitstyle{ergs}} 
\newcommand{\ergs}{\erg}
\newcommand{\ergspersecond}{\erg\unitspace\power{\second}{-1}}

\newcommand{\eV}{\unitstyle{eV}}        
\newcommand{\keV}{\kilo\eV} 
\newcommand{\MeV}{\Mega\eV} 

\newcommand{\Msun}{\ensuremath{M_\odot}}

\newcommand{\parsec}{\unitstyle{pc}}
\newcommand{\kpc}{\kilo\parsec} 

\newcommand{\yr}{\unitstyle{yr}}        
\newcommand{\km}{\kilo\meter}   

\newcommand{\kB}{\ensuremath{k_\mathrm{B}}} 
\newcommand{\pF}{\ensuremath{p_\mathrm{F}}} 
\newcommand{\NA}{\ensuremath{N_\mathrm{\!A}}} 
\newcommand{\mb}{\ensuremath{m_\mathrm{u}}} 

\newcommand{\satellite}[1]{\emph{#1}}

\newcommand{\beppo}{\satellite{BeppoSAX}}
\newcommand{\chandra}{\satellite{Chandra}}

\newcommand{\code}[1]{\textsc{#1}}

\newcommand{\nonsmoker}{\code{non-smoker}}

\newcommand{\lD}{\ensuremath{\lambda_\mathrm{D}}}
\newcommand{\iso}[2]{\ensuremath{\mathrm{^{#1}#2}}}

\newcommand{\Zbar}[1][]{\ensuremath{\langle Z^{#1}\rangle}}
\newcommand{\Abar}{\ensuremath{\langle A\rangle}}
\newcommand{\wpl}{\ensuremath{\omega_{\mathrm{p}}}}

\newcommand{\Leimp}{\ensuremath{\Lambda_{e,\mathrm{imp}}}}
\newcommand{\mdot}{\ensuremath{\dot{m}}}
\newcommand{\medd}{\ensuremath{\mdot_{\mathrm{Edd}}}}

\newcommand{\vsed}{\ensuremath{w_{\mathrm{sed}}}}
\newcommand{\lsed}{\ensuremath{L_\mathrm{sed}}}
\newcommand{\CP}{\ensuremath{C_P}}
\newcommand{\enuc}{\ensuremath{\varepsilon_\mathrm{nucl}}}
\newcommand{\ecool}{\ensuremath{\varepsilon_\mathrm{cool}}}
\newcommand{\gpscps}{\ensuremath{\gram\nsp\cm^{-2}\nsp\second^{-1}}}
\newcommand{\source}[3]{#1~#2$#3$} 
\newcommand{\ycross}{\ensuremath{y_{\mathrm{cross}}}}
\newcommand{\mdotc}{\ensuremath{\mdot_{\mathrm{crit}}}}
\newcommand{\ee}[1]{\ensuremath{\times 10^{#1}}}
\newcommand{\yturn}{\ensuremath{y_{\mathrm{turn}}}}

\begin{document}

\title{Sedimentation and Type I X-ray Bursts at Low Accretion Rates}

\author{Fang Peng\altaffilmark{1}, Edward F. Brown\altaffilmark{2}, and James W. Truran\altaffilmark{1,3}}
\altaffiltext{1}{Department of Astronomy \& Astrophysics, the Center for Astrophysical Thermonuclear Flashes, and the Joint Institute for Nuclear Astrophysics, University of Chicago, Chicago, IL 60637}
\altaffiltext{2}{Department of Physics \& Astronomy, National Superconducting Cyclotron Laboratory, and the Joint Institute for Nuclear Astrophysics, Michigan State University, East Lansing, MI 48824}
\altaffiltext{3}{Argonne National Laboratory,  Argonne, IL 60439}
\email{fpeng@oddjob.uchicago.edu, ebrown@pa.msu.edu, truran@nova.uchicago.edu}

\begin{abstract}
Neutron stars, with their strong surface gravity, have interestingly short timescales for the sedimentation of heavy elements.  Motivated by observations of Type I X-ray bursts from sources with extremely low persistent accretion luminosities, $L_X < 10^{36}\usp\ergspersecond (\simeq 0.01\ensuremath{L_{\mathrm{Edd}}}$), we study how sedimentation affects the distribution of isotopes and the ignition of H and He in the envelope of an accreting neutron star. For 
local mass accretion rates  $\mdot \lesssim 10^{-2}\medd$
(for which the ignition of H is unstable), where $\medd = 8.8\times 10^{4}\nsp\gpscps$, the helium and CNO elements sediment out of the accreted fuel before reaching a temperature where H would ignite. Using one-zone calculations of the thermonuclear burning, 
we find a range of accretion rates  for which the unstable H ignition does not trigger unstable He burning.  This range depends on the emergent flux from reactions in the deep neutron star crust; for $F = 0.1\nsp\MeV(\dot{m}/\mb)$, the range is $3\times 10^{-3}\medd\lesssim\mdot\lesssim 10^{-2}\medd$. 
We speculate that sources accreting in this range will build up a massive He layer that later produces an energetic and long X-ray burst. At  mass accretion rates lower than this range, we find that the H flash leads to a strong mixed H/He flash. Surprisingly, even at accretion rates $\mdot \gtrsim 0.1\medd$, although the H and He do not completely segregate, the H abundance at the base of the accumulated layer is still reduced. While following the evolution of the X-ray burst is beyond the scope of this introductory paper, we note that the reduced proton-to-seed ratio favors the production of \iso{12}{C}---an important ingredient for subsequent superbursts.
\end{abstract}

\keywords{diffusion --- stars: neutron --- X-rays: binaries --- X-rays: bursts}

\section{Introduction}\label{introduction}

An ionized plasma in a gravitational field develops an electric field sufficient to levitate the ions and ensure overall charge neutrality. When there is more than one species of ion present, the ions will experience a differential force: lighter ions float upward (defined by the local gravitational field) and heavier ions sink downward. The high surface gravity
of compact objects makes the timescale for isotopes to stratify often
comparable to or faster than other timescales of interest. In particular, sedimentation of heavier isotopes\footnote{Throughout this paper we use the term sedimentation to describe the separation between species due to an external field. Diffusion in this context refers to the collision-mediated transport driven by abundance gradients.} is important in understanding the surface compositions of white dwarf stars \citep{Vauclair1979The-chemical-ev,paquette86} and isolated neutron stars \citep{chang.bildsten:diffusive}. For accreting white dwarfs, diffusion between the accreted envelope and underlying white dwarf was proposed as a means to enrich
the ejecta of classical novae in CNO isotopes \citep*{prialnik.kovetz:effect,Iben1992Diffusion-and-m}.

Accreting neutron stars, with their strong surface gravity $\approx 2.0\times
10^{14}\nsp\cm\usp\second^{-2}$, are an ideal place to look for the effects of sedimentation.  The sedimentation of heavy elements and resulting nucleosynthesis in the envelope of isolated neutron stars cooling from birth was first described by \citet{rosen:nucleosynthesis,rosen:particle} and has been studied in detail by \citet{chang.bildsten:diffusive,chang.bildsten:evolution}. For accreting neutron stars, the rapid stratification removes heavy nuclei from the photosphere for accretion rates $\dot{M} \lesssim 10^{-12}\Msun\usp\yr^{-1}$ \citep*{bildsten92}. Deeper in the neutron star envelope, the differentiation of the isotopes can alter the nuclear burning, namely unstable H/He burning and rapid proton-capture process (rp-process), that powers Type I X-ray bursts. Some estimates of the relative importance of sedimentation and diffusion were made by \citet*{wallace.woosley.ea:thermonuclear}, who studied accretion at $\dot{M}=2\times 10^{-11}\nsp\Msun\usp\yr^{-1}$ and argued that the partial separation of the H and He layers might play a role in setting the ignition conditions and subsequent burst nucleosynthesis.

Recent long-term monitoring of the galactic center with \beppo\ led to the discovery of nine ``burst-only sources'' \citep[see][and references therein]{Cornelisse2004Burst-only-sour}: \source{SAX}{J1324.5}{-6313}, \source{1RXS }{J1718.4{-4029}, \source{GRS}{1741.9}{-2853}, \source{SAX}{J1752.3}{-3138}, \source{SAX}{J1743.5}{-2349}, \source{SAX}{J1806.5}{-2215}, \source{SAX} J1818.7}{+1424}, \source{SAX}{J1828.5}{-1037},  and \source{SAX}{J2224.9}{+5421}. These sources did not have persistent fluxes detectable with the \beppo/WFC, thus must have extremely low accretion luminosities ($L_X < 10^{36}\usp\ergspersecond \simeq 0.01\ensuremath{L_{\mathrm{Edd}}}$). Even earlier, \citet{Gotthelf1997An-Unusual-X-Ra} had detected a low-luminosity X-ray burst (peak burst luminosity $\approx 0.02$ \ensuremath{L_{\mathrm{Edd}}}) from the globular cluster M28. If the accretion is not concentrated onto a small surface area, so that the local accretion rate is $\mdot\approx \dot{M}/(4\pi R^{2})$, then the sedimentation timescale for CNO nuclei, defined as the time required to move a scale height relative to the center of mass of a fluid element, is less than the accretion timescale, defined as the time for a fluid element to reach a given depth. This is demonstrated in  \S~\ref{diffusion}, where we show that the sedimentation timescale of CNO nuclei is less than the accretion timescale at $\mdot < 0.05\,\medd$. Here $\medd = 2 m_{\rm p} c/[(1+X_{\rm H}) \sigma_{\rm TH} R] = 8.8\times 10^{4}\nsp\gpscps$ is the local Eddington mass accretion rate for a solar composition,
$\sigma_{\rm TH}$ is the Thomson scattering cross section, and $X_{\rm H}$ is the hydrogen mass fraction. 
An estimate of this accretion rate was noted earlier by \citet{wallace.woosley.ea:thermonuclear}. As a result, sedimentation in the accreted envelope must be considered in treating the unstable ignition of hydrogen and helium for these low-$\mdot$ sources. 

Motivated by these observations, we explore in this paper the unstable ignition of hydrogen and helium at low accretion rates and pay particular attention to the regime where the mass accretion rate is less than the critical rate needed for stable H burning \citep*{fujimoto81:_shell_x}. We study how the sedimentation of CNO nuclei affects the unstable ignition of H and He in an accreted neutron star envelope.  At low accretion rates, for which the sedimentation timescale is comparable to or shorter than the burst recurrence timescale, the partial separation of the H and He layers is important in setting the ignition conditions and subsequent burst nucleosynthesis. In particular, this affects the issue of whether unstable H burning triggers the X-ray burst. 
We find that there is a range of mass accretion rates for which H burning is unstable (according to a simple one-zone prescription) but too weak to ignite He. We speculate that this may allow the formation of a deep, cold He layer, and we discuss how the outcome depends on the flux emergent from reactions in the deep crust of the neutron star. At accretion rates lower than this range, we find that the H flash leads to a strong H/He flash. Our work is directly applicable to the bursts observed from the ``burst-only sources'' \citep{Cornelisse2004Burst-only-sour}, for which the time between bursts is observed to be longer than the sedimentation timescale. 

At higher accretion rates, sedimentation is too slow to separate the accreted envelope; nevertheless, an abundance gradient develops. The concentration of H at the base of the accreted envelope is reduced, both because H tends to rise and CNO to sink and also because the H is consumed faster due to the enhanced CNO abundance. Thus, at the depth of burst ignition (triggered by the $3\alpha$ reaction) the proton abundance is reduced. This may be important for determining the amount of \iso{12}{C} produced, which is the most likely fuel for igniting superbursts \citep{cumming.bildsten:carbon}. \emph{We find that the elemental  separation plays a role in setting the ignition conditions of the  X-ray burst and the resulting nucleosynthesis even at higher accretion rates $\dot{m} > 0.1 \medd$.}

In this paper, we first provide, in \S~\ref{diffusion}, a review of the formalism for calculating the elemental sedimentation. We then incorporate, in \S~\ref{model}, our calculations of the differential isotopic velocities into a model of the accumulating neutron star envelope.  With this model we examine (\S~\ref{sec:Results}) the relative abundances of H, He, and CNO isotopes at ignition,  and we use a one-zone model to follow the subsequent thermonuclear burning.  We conclude in \S~\ref{sec:disc-concl} by discussing directions for future research.

\section{Sedimentation and Diffusion}\label{diffusion} 

To describe a multifluid gas, we follow the treatment of  \citet{burgers69:composite_gases}, which constructs the equations from successive moments of the Boltzmann equation. Our plane-parallel atmosphere is in hydrostatic equilibrium, and for simplicity we neglect the terms coupling thermal and particle diffusion \citep[see][and references
therein]{paquette.pelletier.ea:diffusion}.  From these assumptions, we then
have for each species $s$ an equation of continuity and momentum conservation,
\begin{eqnarray}
  \frac{D_s}{Dt}n_s + n_s \frac{\partial u_s}{\partial r} &=& 0~
  \label{continuity.e}\ ,\\
  \frac{\partial P_s}{\partial r} + n_s A_s \mb g - n_s Z_s eE &=& \sum_t K_{st} (w_t - w_s).
  \label{berger.e}
\end{eqnarray}
Here species $s$ has  mass $A_s \mb$, \mb\ being the atomic mass unit, charge $Z_s e$,
density $n_s$, partial pressure $P_s$, and velocity $u_s$. We denote the substantial
derivative for this species by $D_s/Dt \equiv \partial_t + u_s \partial_r$. The force terms in equation~(\ref{berger.e}) result from the gravitational field $g$ and the induced electric field $E$. The center-of-mass of a fluid element moves with a velocity $u = \sum_s n_s A_s u_s/\sum_s n_s A_s$, and $u = \mdot/\rho$ to a good approximation \citep{bildsten92,brown98}. The differential, or diffusion, velocity between species $s$ and the fluid element is then $w_s = u_s - u$. These diffusion velocities satisfy mass and charge conservation,
\begin{eqnarray}
  \label{mc.e}
  \sum_s A_s n_s w_s &=& 0, \\
  \label{cc.e}
  \sum_s Z_s n_s w_s &=& 0,
\end{eqnarray}
with the electrons included in the summation.

The right-hand side of equation~(\ref{berger.e}) is the collision term between species $s$ and $t$, with the resistance coefficient $K_{st} = n_s n_t \langle \sigma_{st} v_{st} \rangle$ being the velocity-weighted cross-section and $v_{st}$ the center-of-mass relative velocity between particles of types $s$ and $t$. The nature of the ionic interaction is characterized by
\begin{equation}
\label{e.Gamma}
\Gamma = \frac{\Zbar[2] e^2}{a\kB T} = 0.11\frac{\Zbar[2]}{\Abar^{1/3}}
\left(\frac{\rho}{10^{5}\nsp\grampercc}\right)^{1/3} \left(\frac{10^{8}\nsp\K}{T}\right) \ ,
\end{equation}
where $a$ is the average ion spacing, defined by $4 \pi a^3 n/3 = 1$ with $n = \rho\NA/\Abar$.
The ions are not strongly interacting, and the temperature is sufficiently high that the ion quantum occupation is small, $n/n_{\mathrm{Q}} = n[2\pi\hbar^{2}/(A \mb\kB T)]^{3/2} = 3.2\times 10^{-4}A^{-5/2}(\rho/10^{5}\nsp\grampercc)(T/10^{8}\nsp\K)^{3/2}\ll 1$. The ions are not as strongly interacting as in cooling white dwarfs \citep[see][and references therein]{deloye.bildsten:gravitational}. The unstable ignition of H and He occurs where $\Gamma \lesssim 1$, which is where the resistance coefficients $K$ are most uncertain. Appendix \ref{resistance} describes in detail our choice for computing the resistance coefficients; in short, we use the fit of \citet{fontaine.michaud:diffusion}, which agrees with other fits \citep{muchmore:diffusion,paquette.pelletier.ea:diffusion} in the regime $\Gamma \sim 1$, has the correct scaling at $\Gamma \gg 1$ \citep*{bildsten.hall:diffusion,hansen.joly.ea:self-diffusion}, and goes, in the limit $\Gamma\ll 1$, to $K_{st}^{0}$ (see Appendix~\ref{resistance}) computed using a Coulomb potential with a long-range cutoff \citep{chapman.cowling,burgers69:composite_gases}.

Before describing our detailed calculations of the accreted envelope structure, we present a simpler case to illustrate how sedimentation changes the structure of the envelope.
In the case of a trace species, labeled 2, in a background of a species labeled 1, i.e., $n_1 \gg
n_2$ and $w_1 = 0$, equations~(\ref{berger.e}) separate: the right-hand side of the equations vanishes, thereby fixing the electric field, $eE = \mb g A_1/(Z_1 + 1)$
where the electrons are non-degenerate and $eE = \mb g A_1/Z_1$ where
the electrons are degenerate.  Substituting $E$ into the
equation of motion for species 2 then determines the sedimentation velocity,
$\vsed = w_2 = n_2 (A_2 \mb g - Z_2 eE)/K_{12}$.  We chose the
sign of $\vsed$ to be positive if species 2 moves downward.

Using the Stokes-Einstein relation to determine $K_{12}$ from the drag
coefficient for a liquid sphere (\citealt{bildsten.hall:diffusion}; see Appendix~\ref{resistance}), and a nonrelativistic electron equation of state, we find the sedimentation velocity of a trace nucleus 
\citep*[see][]{brown.bildsten.ea:variability} 
\begin{equation}\label{e.vsed}
\vsed = 2 \times 10^{-3} g_{14}\frac{T_7^{0.3}}{\rho_5^{0.6}}
\frac{(A_2Z_1 - Z_2A_1)A_1^{0.1}}{Z_1^{2.3}Z_2^{0.3}}\nsp\cm\usp\second^{-1}, 
\end{equation}
where we use the common shorthand 
$g_{14} = g/(10^{14}\nsp\cm\usp\second^{-2})$, $T_7 = T/(10^7\usp\K)$, and $\rho_5 = \rho/(10^5\usp\grampercc)$.
In a pure H plasma, the sedimentation velocity\footnote{Throughout this paper we use a
  Newtonian metric and assume a neutron star mass $1.4\nsp\Msun$ and
  radius 10\nsp\kilo\meter, so that $g_{14} = 1.9$.} is greater than the mean velocity $u = \mdot/\rho$ for 
$\mdot < 400 \usp\gpscps T_7^{0.3} \rho_5^{0.4} (A_2 - Z_2)/Z_2^{0.3}$; under
conditions at which H ignites ($T_7 \sim 5; \rho_5 \sim 4$) 
this corresponds to $\mdot < 0.02\medd$ and
$\mdot < 0.05 \medd$ for a trace \iso{4}He and
\iso{12}C nucleus, respectively.

\begin{figure}[t]
  \centering
  \includegraphics[width=3.4in]{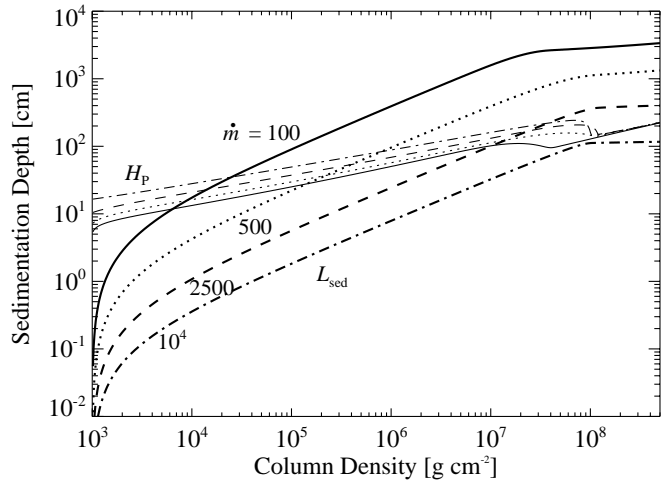}
  \caption{Sedimentation depth $\lsed$ for \iso{4}{He} (\emph{thick lines}) and pressure scale height $H_{p}$ (\emph{thin lines}) as a function of
    column density for 4 mass accretion rates: $\mdot = 100$ (\emph{solid lines}), 500 (\emph{dotted lines}), 2500 (\emph{dashed lines}) and $10^{4}\nsp\gpscps$ (\emph{dot-dashed lines}).}
  \label{sedi.f}
\end{figure}

For a more refined estimate, we transform to a frame co-moving with a fluid element and solve for the displacement of a test particle from the center-of-mass of the fluid element,
\begin{equation}
  \label{eq:lsed}
  \frac{d\lsed}{dy} = \frac{d\lsed}{dt}\mdot^{-1} = \vsed \mdot^{-1},
\end{equation}
while simultaneously solving for the thermal and compositional structure
of the accreting envelope (see \S~\ref{model}, eq.~[\ref{radiation.e}] and [\ref{heat.e}]).  
Here $dy = - \rho dz = \mdot dt$ is the column density.
The electric field is
computed using an analytical formula \citep{chang.bildsten:diffusive} that is valid for arbitrary electron degeneracy. Figure~\ref{sedi.f} shows \lsed\ (\emph{thick lines}) for a trace \iso{4}{He} nucleus in a bath of H as a function of the column $y$. For comparison, we also plot the pressure scale height $H_P = P/(\rho g)$ (\emph{thin lines}). Both \lsed\ and $H_{P}$ are shown for 4 different values of \mdot: 100 (\emph{solid lines}), 500 (\emph{dotted lines}), 2500 (\emph{dashed lines}) and $10^4\nsp\gpscps$ (\emph{dot-dashed line}).  For later convenience, we define $y_{\mathrm{cross}}(\mdot)$ to be the column density at which $\lsed = H_P$. An accreted fluid element will be significantly stratified by the time it reaches a column \ycross.  For a fluid mixture, the growing concentration gradient slows the drift velocity as diffusion becomes important.  The balance of diffusion and sedimentation sets the equilibrium abundance profile, and so over the time to reach a column \ycross\ the fluid element roughly comes into diffusive equilibrium.
$y_{\mathrm{cross}}(\mdot)$ is a strong function of accretion rate and surface gravity: for a solar composition plasma and the velocity estimate for a trace He nucleus in H (eq.~[\ref{e.vsed}])
and the scale height for a degenerate, nonrelativistic electron gas, $\ycross \approx 2.4\times 10^{8} \mdot_{3}^{25/6} T_{7}^{-5/4}g_{14}^{-31/6}\nsp\gram\usp\cm^{-2}$.  For trace CNO elements in H, $\ycross \approx 10^{7}\mdot_{3}^{25/6} T_{7}^{-5/4}g_{14}^{-31/6}\nsp\gram\usp\cm^{-2}$. Note that with the approximations in eq.~(\ref{e.vsed}), the sedimentation velocity of trace CNO elements in He is zero and we must use the full equation of state, including Coulomb effects, to compute \vsed.

For HCNO burning, H is consumed on a timescale $\approx 5 \times 10^4
\nsp(0.02/Z_\mathrm{CNO})(X_\mathrm{H}/0.71)\nsp\second$, or equivalently at a column
\begin{equation}
y_{\mathrm{burn}}^{\mathrm{H}} = 
-\mdot\left(\frac{X_{\mathrm{H}}}{\dot{X}_{\mathrm{H}}}\right) 
= 3 \times 10^8\nsp{\rm g\,cm^{-2}} \left(\frac{\dot{m}}{0.1\nsp\dot{m}_{\rm Edd}}\right)
\left(\frac{0.02}{Z_\mathrm{CNO}}\right) \left(\frac{X_\mathrm{H}}{0.71}\right) .
\end{equation}
Using the previously defined $\ycross$,  we determine the mass accretion rate below which the CNO elements are separated from H by setting $\ycross = y_{\mathrm{burn}}^{\mathrm{H}}$. 
We find the critical accretion rate at which the CNO nuclei and H nuclei separate on a timescale comparable to them to ignite, $\mdotc/\medd \approx 0.02 T_{7}^{15/38} g_{14}^{31/19}$.  For typical conditions for H ignition, $T_{7}\approx 5$, and using $g_{14}=1.9$, we estimate that CNO elements are depleted before H ignites for $\mdot \lesssim 0.1\medd$.
Helium sinks more slowly than CNO; in the absence
of CNO sedimentation, He would settle out of the H layer prior to H
ignition for $\mdot \lesssim  0.06~\medd$.  These simple estimates will be a useful guide for our more detailed solutions, as we now describe.

\section{Evolution of an Accreting Envelope with Sedimentation}\label{model}

We now build on the simple estimates of section~\ref{diffusion} and numerically solve for evolution of an accreting envelope prior to unstable ignition. In addition to equations (\ref{continuity.e}) and (\ref{berger.e}) we have the equations for temperature and heat flux, 
\begin{eqnarray}
  \label{radiation.e}
  \frac{\partial T}{\partial y} &=& \frac{F}{\rho K}\ , \\
  \label{heat.e}
  \frac{\partial F}{\partial y} &=& 
  \CP\left(\frac{\partial T}{\partial t} + 
    \mdot\frac{\partial{T}}{\partial{y}}\right)
  - \frac{\CP T\mdot}{y}\nabla_\mathrm{ad} - \enuc\ ,
\end{eqnarray}
where \CP\ is the specific heat at constant pressure and
$\nabla_\mathrm{ad} \equiv (\partial\ln T/\partial\ln P)_S$.  Note that we have neglected heat transport carried by the rising and sinking isotopes, as heat transport by photons and semi-degenerate electrons will be much more efficient.  The compressional flux (terms proportional to \mdot\ in eq.~[\ref{heat.e}]) are of order $\sim \keV \dot{m}/m_{u}$ at these accretion rates \citep{brown98} and are smaller than the flux from steady H burning (for $\dot{m}\gtrsim 0.01\medd$) or from electron captures in the crust ($\gtrsim 0.1\nsp\MeV \mdot/m_u$; \citealp{brown:nuclear}).

We use a radiative-zero outer boundary, $F = 4/(3 \kappa y)\left[\sigma_{\rm R} T^4 - 
F_{\rm acc} \right]$, where $F_{\rm acc} \equiv GM\mdot/(2R)$ is the accretion flux and $\kappa(\rho[y,T],T)$ is the opacity.
For the inner boundary, at the bottom of the ash layer, we set the flux to a constant value.  This flux is set by reactions in the deep crust---electron captures, neutron emissions, and pycnonuclear reactions \citep{sato79,haensel90a,haensel.zdunik:nuclear}---and for $\mdot \gtrsim 0.1\nsp\medd$ is roughly $F  = 0.1\nsp\MeV \nsp\mdot/\mb$ \citep{brown:nuclear}; at lower accretion rates the heat per nucleon flowing out is increased. Here we have computed solutions for different values of $F\mb/\mdot$ (see \S~\ref{sec:Results}).

To simplify our calculations, we use the disparity between the thermal and accretion timescales, $\CP y^2/(\rho K) \ll y/\mdot$. At each timestep we solve equations~(\ref{radiation.e}) and (\ref{heat.e}) with $\partial T/\partial t \to 0$, we presume that the envelope evolves through a sequence of steady-state thermal profiles. This technique has been  used in previous studies \citep{cumming.bildsten:rotational,Cooper2005On-the-Producti} of H/He ignition.

\subsection{Microphysics of the Fuel Layer: Equation of State, Thermal Conductivity, and Nuclear Reactions}\label{eos}

Closing equations (\ref{continuity.e})--(\ref{cc.e}),  (\ref{radiation.e}), and (\ref{heat.e}) requires a specification of the equation of state, the thermal conductivity, and the nuclear heating rate \enuc. For the thermal properties, we mostly follow previous papers \citep{brown:nuclear,brown.bildsten.ea:variability}, and here shall just review our choices. We use a tabulation of the Helmholtz free energy \citep{timmes.swesty:accuracy} to compute the electronic EOS. We compute the ionic free energy for the liquid phase, $1\le\Gamma < 175$, from the fit of \citet{chabrier98} and for the solid phase from the fit of \citet{farouki93}. The free energies of these phases are equal at $\Gamma = 178$. We assume throughout that the ions are classical, i.e., we ignore quantization of phonon modes.  For the temperatures and densities of interest, this is a good approximation.

The relevant opacities for the temperatures and densities in the neutron star envelope are Thomson scattering and free-free absorption.  For free-free opacity, we use the fit from \citet{schatz99} which is reasonably accurate (fractional errors $\sim 10\usp\%$) when compared against the calculations of \citet*{itoh85:_relat,itoh91:_rossel_gaunt}.  We calculate the Thomson scattering opacity by using a fit \citep{buchler.yueh:compton} that reproduces the non-degenerate limit \citep{sampson:opacity} and includes corrections for the relativistic and degenerate electrons. For the electron thermal conductivity, we write the 
relaxation time as
$\tau^{-1} = \tau_\mathrm{e,e}^{-1} + \tau_\mathrm{e,ion}^{-1}$.  Here the term for electron-electron scattering, $\tau_{\mathrm{e,e}}$ is computed from the fit of \citet{potekhin97} to the results of \citet{urpin80:_therm}.  We write the relaxation time for electron-ion scattering as
\begin{equation}
   \tau_{e,\mathrm{ion}}^{-1} = \frac{4\pi e^{4}}{\pF^{2}v_\mathrm{F}}
   \frac{\rho}{\Abar\mb} \left(\Zbar^2 \bar{\Lambda}_{e,\mathrm{ion}} +
   Q\Leimp \right)\ .  \label{eq:tau-mi}
\end{equation}
Here $\bar{\Lambda}_{e,\mathrm{ion}}$ is the Coulomb logarithm for a
single ion of charge number $\Zbar$ and mass number $\Abar$; $Q =
\Zbar[2]-\Zbar^{2}$ measures
the impurity concentration; and \Leimp\ is the Coulomb logarithm
computed under the assumption that the ions are randomly distributed
\citep{itoh93}.  For $\bar{\Lambda}_{e,\mathrm{ion}}$ we use the fitting formula of \citet{potekhin99:_trans}. 

To follow the nuclear burning, we use a reaction network with 22 isotopes: \iso{1-2}{H}, \iso{3-4}{He}, \iso{7}{Li}, \iso{7}{Be}, \iso{8}{B}, \iso{12-13}{C}, \iso{13-15}{N}, \iso{14-18}{O}, \iso{17-19}{F}, and \iso{18-19}{Ne}. The light isotopes are included to cover hydrogen burning through p-p chains that are 
important at the low temperature end ($\mdot <   5.7\ee{-4}\medd = 50\nsp\gpscps $).
The reaction rates are taken from the compilation REACLIB 
\citep[see][and references therein]{thielemann86,rauscher00}, and most of them are originally from the compilation of \citet{caughlan88:_therm}.
These rates are formally fitted for $T > 10^8\usp\K$, but have the correct asymptotic form for 
the low-temperature reactions of interest here.  We neglect electron screening.  The rates for which experimental data were not available are typically taken from shell-model calculations \citep{rauscher00}. We include weak rates from \citet{Fuller1982Stellar-weak-in} and from \citet{langanke.martinez-pinedo:weak}; weak interactions are not important, however, under conditions of interest in this paper.

\subsection{Numerical Solution for the Diffusive Velocities}\label{numerical}

For a mixture of $N$ ion species there are $N+2$ unknowns, the
$N+1$ velocities $w_s$ of the ions and electrons and the electric field $E$; for these unknowns
there are $N+2$ equations (eqs.~[\ref{berger.e}]--[\ref{cc.e}]).
Inspection of equation~(\ref{berger.e}) reveals the velocities $w_s$ to
contain an advective part depending on $g$, $E$, and $\partial_r T$, and
a diffusive part depending on $\partial_r n_s$.  The mixed advective-diffusive nature of the equations makes simple centered-difference schemes, such as Crank-Nicholson, unstable \citep[see, for example,][]{Ferziger2002Computational-M}.  For numerical
stability, we adopt a technique implemented by
\citet{iben.macdonald:effects} and used in studies of the hydrogen-shell
flash in evolved stars \citep*{althaus.serenelli.ea:diffusion} and
pulsational instabilities in white dwarf stars
\citep{gautschy.althaus:nonadiabatic,althaus.c-orsico:double-layered}.
The ion and electron diffusion velocities and the electric field are separated into two pieces,
\begin{eqnarray}
  \label{wi.e}
  w_i &=& w_i^g - \sum_{j=\mathrm{ions}} \sigma_{ij} \frac{d\ln
    n_j}{dr}\ , \\
  \label{eE.e}
  E &=& E^g - \sum_{j=\mathrm{ions}} \beta_j \frac{d\ln n_j}{dr}\ ,
\end{eqnarray}
where $w^g$ and $E^g$ are the components due to the gravitational and
electrical forces, and the sums on the right-hand side are only over the ions.  
Throughout this paper, we
ignore the effect of thermal diffusion, as this term is negligible under
the conditions of interest
\citep{stevenson.salpeter:phase,muchmore:diffusion}. Inserting
equations (\ref{wi.e}) and (\ref{eE.e}) into equations
(\ref{berger.e})--(\ref{cc.e}), we use the fact that these equations must hold for arbitrary number density gradients for each species to equate the coefficients of $d\ln n_{j}/dr |_{j=1,N}$. This gives us $N(N+2)$ equations for the variables $(\sigma_{ij},\beta_j)$, in addition to the equations for the  $N+2$ variables $(w_s^g,E^g)$. We solve these linear equations by direct matrix inversion.

With the $\sigma_{ij}$ and $\beta_j$ determined, we insert
equation~(\ref{wi.e}) into the isotopic continuity equation
(eq.~[\ref{continuity.e}]) and obtain
\begin{eqnarray}
  \frac{DY_i}{Dt} &=& \frac{\partial}{\partial y}\left(\rho w_i Y_i\right) 
  \nonumber \\
  &=& \frac{\partial}{\partial y}\left[\rho Y_i 
    \left(w_i^g - \sum_j \sigma_{ij} \frac{d\ln n_j}{dr}\right)\right]\ ,
  \label{continuityLag.e}
\end{eqnarray} 
where $Y_i = n_i/(\rho \NA)$ and $y\equiv \int_{z}^{\infty}\rho dz$ is
the column density.  We solve for the $Y_i$ in each spatial zone with a
semi-implicit finite difference method \citep{iben.macdonald:effects}.  
We tested this scheme by defining a closed box with an initially uniform mixture of H/He and an isothermal temperature profile. We evolved the box in time until it reached diffusive equilibrium, and the resulting chemical profile was in good agreement with the analytical result of \citet[appendix]{alcock:diffusion}.\footnote{Note that the terms $\exp(-5x/3)$ in equation (A14) of \citet{alcock:diffusion} should be $\exp(-5x/2)$.} 

\subsection{Evolution to Ignition}\label{sec:evolution-ignition}

We  now implement the Lagrangian expression for the abundance evolution (eq.~[\ref{continuityLag.e}]) in a model of the accumulating neutron star envelope.  We take the
composition of matter accreted from the companion star to be roughly solar composition in \iso{1}{H} and \iso{4}{He}, and distribute the remaining mass evenly between
\iso{12}{C} and \iso{16}{O}:
$X(\iso{1}{H})=0.71$, $X(\iso{4}{He})=0.27$, $X(\iso{12}{C}) = 0.01$, 
and $X(\iso{16}{O}) = 0.01$.  Below the fuel layer are ashes
from previous X-ray bursts; we set the ash composition to \iso{64}{Zn}, consistent with the findings of recent one-dimensional calculations of repeated X-ray bursts \citep{woosley.heger.ea:models,Fisker2005Extracting-the-} that find $A\approx 60$ to be favored.  We do not follow mixing between the fuel and ash layers and therefore this simple model neglects effects arising from  ``compositional inertia'' \citep{Taam1993Successive-X-ra}.

For simplicity our scheme employs a fixed number of Lagrangian zones, so that our grid follows the advected fluid elements. A computation domain is closed, i.e., we set the diffusion velocities to zero at the boundaries. Our scheme is not as refined as schemes that add zones to the computational domain. The error added to this scheme is that H tends to accumulate at the upper boundary, which inhibits further sedimentation. Our results therefore err in the direction of a lesser separation of H and CNO, which is adequate for our initial exploration.

We determine the size of a timestep by choosing the minimum of the thermal timescale, $C_py^2/(\rho K)$,  the sedimentation timescale, $\Delta y/(\rho w_s)$, and the burning timescale ($|Y/\dot{Y}|_{\mathrm{nuc}}$). 
At each timestep we advect the matter inward and increase the pressure in each zone. We then update the composition using the reaction network followed by a diffusion step (eq.~[\ref{wi.e}]--[\ref{continuityLag.e}]).  Finally, we update the thermal profile by solving  equations (\ref{radiation.e}) and (\ref{heat.e}) via relaxation  \citep{pre92}. This process is repeated at each timestep until a thermal instability develops.
We determine this ignition point using a simple one-zone  criterion,
$(\partial\enuc/\partial T)_P > (\partial \ecool/\partial T)_P$, where 
\enuc\ is the nuclear heating rate and $\ecool =
\rho KT/y^2$ is an approximation of the local cooling rate. Here 
$K$ is the thermal conductivity. This scheme quantitatively agrees with linear stability analyses  on ignition column densities at regions where both methods find instability
\citep{narayan.heyl:thermonuclear,Cooper2005On-the-Producti}, and it also roughly agrees with the ignition conditions found by multizone calculations \citep{woosley.heger.ea:models,Fisker2005Extracting-the-}.

\section{Results}\label{sec:Results}

Using our accumulation model, we explore the evolution of the accreting neutron
star envelope for accretion rates $10^{-4}\textrm{--}0.6\nsp\medd$ ($10\nsp\gpscps\textrm{--}5\times10^{4}\nsp\gpscps$ for accretion of a solar composition mixture).  For each accretion rate, we ran calculations both with and without isotopic sedimentation, and we follow the thermal and chemical evolution of the mixture until the envelope becomes thermally unstable, as described in \S~\ref{model}.
We first describe the qualitative features of our results (\S~\ref{sec:solutionFeautres}). We then apply our findings to two problems: the phenomenon of X-ray bursts at low accretion rates (\S~\ref{sec:bursts-at-low}), and the production of \iso{12}{C} to fuel subsequent superbursts, appropriate for sources accreting at $\mdot \gtrsim 0.1\medd$ (\S~\ref{sec:carbon-production}).

\subsection{Properties of the Accreted Envelope}\label{sec:solutionFeautres}

\begin{figure}[htb]
\centering
\includegraphics[width=3.4in]{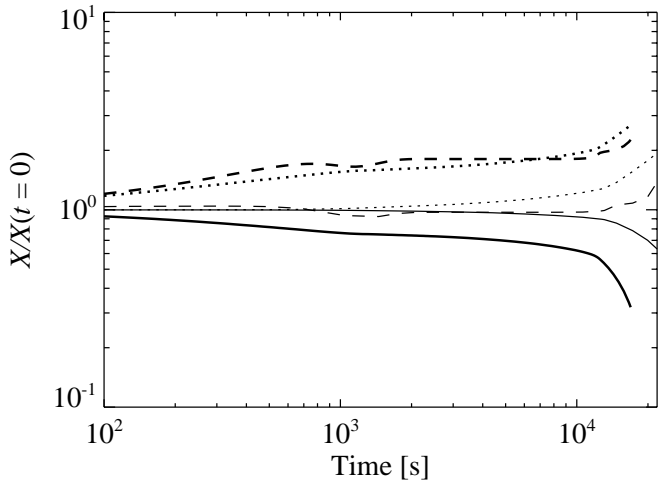}
\caption{The evolution of mass fractions of H (\emph{solid lines}), He (\emph{dotted lines}) and CNO elements (\emph{dashed lines}), normalized to their initial values as a function of Lagrangian time $t = y/\mdot$ for $\mdot = 0.11 \nsp\medd$. We show cases when diffusion is (\emph{thick lines}) and is not (\emph{thin lines}) included. The curves terminate where He ignites unstably.}
  \label{evol2.f}
\end{figure}

At accretion rates for which the temperature is hot enough for hydrogen to burn stably, hydrogen is removed from the base of the accreted fuel both by floating upwards and by capturing onto the enhanced abundance of CNO nuclei that have settled downward.  Figure~\ref{evol2.f} shows the mass fraction of H (\emph{solid lines}), He (\emph{dotted lines}) and CNO nuclei (\emph{dashed lines}), relative to its initial value, at the base of the newly accreted layer for a local mass accretion rate $0.11\nsp\medd$ ($10^{4}\nsp\gpscps$). Hydrogen ignition occurs after $ 10^{3}\nsp\second$.
The mass fraction is plotted against the time since arrival of the fluid element at the photosphere, so that the column corresponding to a given time $t$ is $y(t) = \mdot t$. We show for comparison solutions with diffusion (\emph{thick lines}) and without diffusion (\emph{thin lines}).  At this accretion rate, H is consumed via the HCNO cycle (Fig.~\ref{evol2.f}), which increases the abundance of He. When sedimentation is included, the CNO mass fraction has doubled at the point where the H is consumed by HCNO burning.  As a result, hydrogen is consumed on a shorter timescale, releasing a greater heat flux and increasing the He abundance. At the end of the calculation, the triple alpha reaction has started, increasing the number of CNO seed nuclei. We terminate the curves where the triple alpha reaction becomes thermally unstable; as is clear from the figure, the ignition column decreases by $\approx 20\%$ when sedimentation is included.

\begin{figure}[htb]
\centering
\includegraphics[width=3.4in]{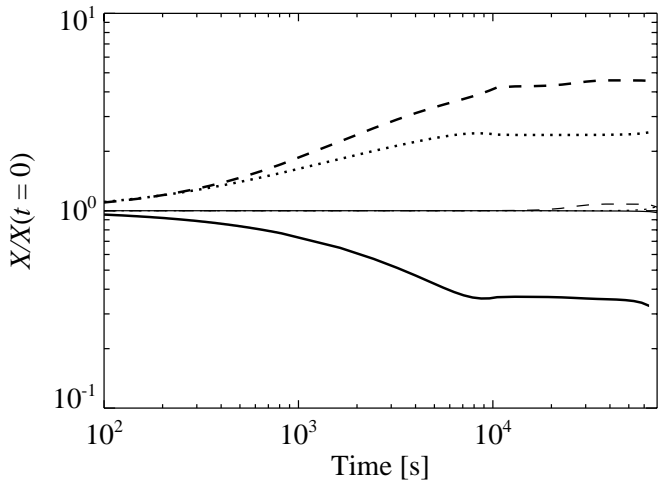}
\caption{Changes in mass fractions at the base of the fuel layer, for $\mdot = 5.7\times10^{-3}\nsp\medd$. The curves have the same meaning as those in Fig.~\protect\ref{evol2.f}.}
\label{evol1.f}
\end{figure}

This effect is even more pronounced at the accretion rate for which H does not burn stably. Figure~\ref{evol1.f} shows the same calculations for a lower accretion rate $\mdot= 5.7\times 10^{-3}\nsp\medd$ ($500\nsp\gpscps$). For this calculation, the base of the accreted envelope reaches diffusive equilibrium in $8\times 10^{3}\nsp\second$ (where the abundance curves flatten). This agrees with our simple estimates from \S~\ref{diffusion} (see Fig.~\ref{sedi.f} and following discussion), namely the time for the envelope to be stratified is $\approx \ycross/\dot{m} \approx 7000\nsp\second$.  Note that as the envelope becomes stratified, the timescale for CNO to sink becomes set by its drift time in a He-rich plasma, which is considerably longer than the timescale for a H-rich plasma (eq.~[\ref{e.vsed}]) and becomes comparable to the timescale for He to sediment. This is why the curves in Fig.~\ref{evol1.f} all change slope at roughly the same time.

We can compare the asymptotic values of the abundances in Figure~\ref{evol1.f} with the analytical diffusive equilibrium solution of \citet[appendix]{alcock:diffusion}.  We use an ideal gas equation of state for simplicity.  For a H/He plasma with $\int X(\iso{1}{H})\,dz = 0.7$ and $\int X(\iso{4}{He})\,dz = 0.3$, the equilibrium solution at the base $z = 0$ has $X(\iso{1}{H}) = 0.29$ and $X(\iso{4}{He}) = 0.71$, ie., the H has decreased to 0.4 of its original abundance.  
For $\iso{12}{C}$ in a H-rich plasma with mass fraction of $X(\iso{1}{H}) = 0.98$ and $X(\iso{12}{C}) = 0.02$, we get $X(\iso{12}{C},z=0) = 0.23$.
If we now approximate the He-rich layer by a box containing He and C with $\int X(\iso{12}{C})\,dz = 0.07$ (so that the ratio of C/He matches 0.02/0.28), the equilibrium solution has $X(\iso{12}{C},z=0) = 0.17$; that is, sedimentation increased the abundance of \iso{12}{C} by a factor 2.4. Our numerical solution has $X(\iso{12}{C},z=0)$ increased by a factor $\approx 4.5$, which is somewhat larger than the analytical solution for \iso{12}{C} in \iso{4}{He}, but much less than that for \iso{12}{C} in H.

\begin{figure}[t]
\centering
\includegraphics[width=3.4in]{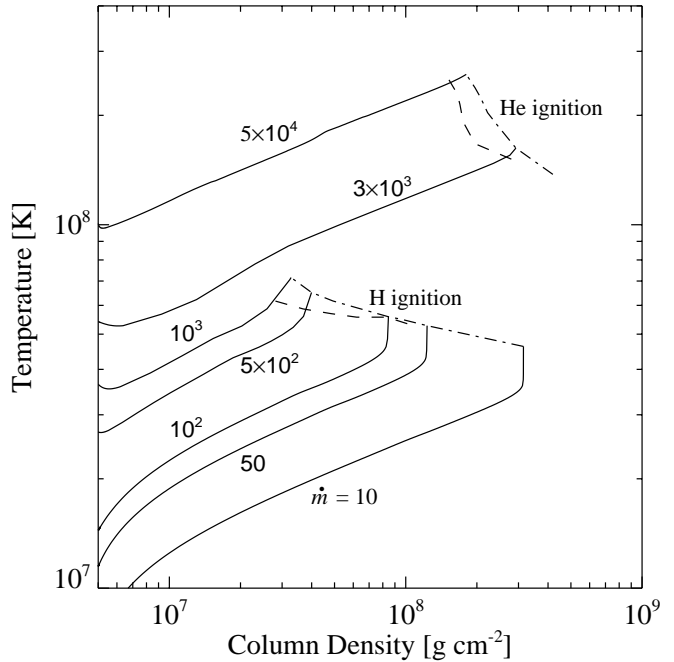}
\caption{
 The temperature evolution of the base of the accreted layer as it is advected to deeper column. The tracks 
(\emph{solid lines}) correspond to different local mass accretion rates and are in units of ${\rm {g\,cm^{-2}\,s^{-1}}}$.  The ignition curves of H and He when sedimentation is  (\emph{dashed lines}) and is not (\emph{dot-dashed lines}) taken into account are shown as well.}  
\label{fuel.f}
\end{figure}

The temperature evolution of our mass shell at the base of the accreted envelope is shown in Figure~\ref{fuel.f}. Each curve is labeled by its local mass accretion rate in units of \gpscps.  The x-axis indicates the column density; the temperature at the base of the  accreted envelope follows the trajectory in $(P=yg,T)$ space shown in the figure. In all of the calculations shown, we include sedimentation. As with Figures~\ref{evol2.f} and \ref{evol1.f}, the curves terminate where the envelope becomes unstable to a thin-shell instability (estimated from a one-zone calculation). The locus of these ignition points are indicated by the curves labeled ``He ignition'' (for $\mdot \geq 2 \times 10^3\nsp\gpscps$). and ``H ignition''. Our conditions for unstable He ignition agree (in the absence of sedimentation and using an emergent flux from the crust $F = 0.15\nsp\MeV (\dot{m}/\mb)$) roughly with those obtained by \citet{cumming.bildsten:rotational}. At $\mdot = 0.1\nsp\medd$, our ignition column is 30\% larger, but our temperature, hydrogen, and helium mass fractions agree to within 7\%. 
This difference is likely caused by the H burning rate being slower than the HCNO limit used by \citet{cumming.bildsten:rotational} at $T_8 < 1.7$. At these temperatures, the reaction $\iso{13}{N}(p,\gamma)\iso{14}{O}$ does not entirely dominate over the $\beta$-decay branch, so that the rate is not entirely set by the decays of \iso{14}{O} and \iso{15}{O}. The longer $\beta$-decay time of \iso{13}{N} ($10~\rm{min}$ halflife) decreases the total rate of H burning from the HCNO limit.

\begin{figure}[b]
\centering
\includegraphics[width=3.4in]{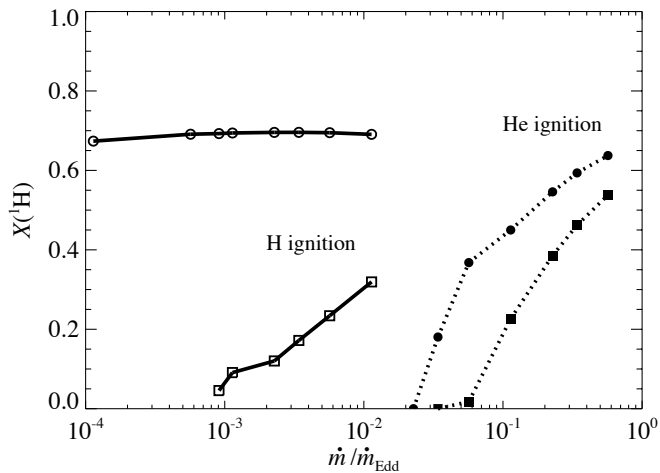}
\caption{
  Mass fraction of hydrogen at the column where either H (\emph{open symbols, solid lines}) or He
  (\emph{filled symbols, dotted lines}) unstably ignites, as a function of mass accretion rates.  We show results for which sedimentation is ignored (\emph{circles}), and for which it is included (\emph{squares}).}
\label{abund_mdot.f}
\end{figure}

As noted by \citet{fujimoto81:_shell_x}, there are three regimes of burning
parameterized by \mdot\ \citep[see][and references
therein]{bildsten:thermonuclear}:
1) $\mdot < \mdot_{c2}$, for which
hydrogen burns unstably;
2) $\mdot_{c2} < \mdot < \mdot_{c1}$, for which
hydrogen burns stably and is completely consumed prior to unstable He
ignition; and 
3) $\mdot > \mdot_{c1}$, for which hydrogen burns stably and
is only partially consumed prior to unstable He ignition. 
In the absence of sedimentation, we find $\mdot_{c2} \approx
10^3\usp\gpscps$, and $\mdot_{c1} \approx 2\times10^3\usp\gpscps$ 
(see Table~\ref{tab:mdot_crit} and Fig.~\ref{abund_mdot.f}). 

\begin{deluxetable}{lll}
\tablewidth{0pc}
\tablecaption{Critical Mass Accretion Rates \label{tab:mdot_crit}}
\tablehead{
\colhead{\rm Reference} & \colhead{$\mdot_{c2}$} & \colhead{$\mdot_{c1}$}  \\
 \colhead{ }                   &  \colhead{($\gpscps$)}  & \colhead{($ \gpscps$)}}
\startdata
this work (no sed.) &  $10^3$    &  $2 \times 10^3$   \\
\citet{hanawa.fujimoto:irreducible}  & $3.2 \times 10^2$ & $3 \times 10^3$ \\
\citet{bildsten:thermonuclear}   & $1.3 \times 10^3$ & $8.6\times 10^3$ \\
\citet{narayan.heyl:thermonuclear} &  $3 \times 10^2$  & $ 3\times10^3 $ \\
this work (with sed.) & $10^3$   &  $5 \times 10^3$   
\enddata
\tablecomments{This comparison is for $M = 1.4\nsp\Msun$, $R=10\nsp\km$, $X=0.7$, and $Z=0.02$. The result of 
\citet{narayan.heyl:thermonuclear} is taken from the case of core temperature $10^8\nsp\K$
and $R=10.4\nsp\km$. For \protect\citet{narayan.heyl:thermonuclear}, we interpret $\mdot_{c2}$ as being the critical mass accretion rate required for prompt hydrogen bursts.}
\end{deluxetable}%

When sedimentation is included, $\mdot_{c2}$ is unchanged:; the abundance of H at the base of the accreted envelope is depressed, however, for  $\mdot < \mdot_{c2}$ (see Fig.~\ref{abund_mdot.f}).  Moreover, for accretion rates $\mdot \lesssim \mdot_{c1} = 5 \times10^{3}\nsp\gpscps$, \emph{He ignites in the absence of H}. We emphasize, however, that the temperature at ignition and the total mass of H in the envelope is only slightly affected by sedimentation. The characteristics of the burst will depend on the interplay between the thermal instability and the growth of the convective zone \citep{woosley.heger.ea:models,Weinberg2005Exposing-the-Nu}; such a study is beyond the scope of this paper, but is clearly a crucial future step for understanding the burst physics.
It is tantalizing that the accretion rate $\mdot_{c1}$ at which mixed H/He ignition occurs is increased by a factor of 2 when sedimentation is taken into account, and we speculate that this might alleviate the discrepancy between the predicted transition in burst duration \citep{fujimoto81:_shell_x} and recent observations \citep[see, for example][]{den-Hartog2003Burst-propertie}. 

\subsection{Bursts at Low Accretion Rates}\label{sec:bursts-at-low}

\subsubsection{Observations}\label{sec:observations-bursts-low}

As discussed in \S~\ref{introduction}, X-ray bursts with extremely low persistent luminosities ($\leq 10^{36}\usp\ergspersecond$) have been discovered recently
\citep[see][]{cocchi.bazzano.ea:discovery-1741,cocchi.bazzano.ea:discovery,kaptein..ea:discovery,cornelisse.verbunt.ea:bepposax,arefiev:type-i}. In Table~\ref{tab:bursts_lowrate}, we list the burst duration and persistent luminosity of all such known burst sources. Several of these did not have persistent fluxes detectable with the \beppo/WFC and are known as ``burst-only sources''.  Follow-up observations with \chandra\ revealed that the sources' persistent luminosities are $10^{32}\textrm{--}10^{33}\usp\ergspersecond$
\citep{cornelisse.verbunt.ea:chandra}, which is consistent with these sources being X-ray transients. 
These bursts are very rare: on average, there is only one burst detected for every $10^{6}\nsp\second$
seconds of \beppo/WFC observation time. The bursts have rather short durations, $\lesssim 20\nsp\second$.  At somewhat higher accretion rates, $\mdot\sim 0.01\medd$, several bursts have been observed with longer durations, $\lesssim 1000\usp\second$, that are intermediate between mixed H/He bursts and superbursts (see the lower part of Table~\ref{tab:bursts_lowrate}). 

We note for completeness that still longer bursts were exhibited by sources at $L\lesssim 0.1 L_\mathrm{Edd}$. \citet{czerny87} observed (with \emph{Einstein}/MPC) a burst of duration $\approx 2500\nsp\second$, when the persistent luminosity was $2.5\times 10^{37}\nsp\ergspersecond (d/5\nsp\kpc)^{2}$ \citep{czerny87}\footnote{We adopt the fiducial distance 5\nsp\kpc\ recommended by \citet{rutledge.ea.01b:aqlx1}}.  \citet{fushiki92} interpreted the burst as due to electron captures at great depth. This burst had an initial e-folding time $\sim 100\nsp\second$, and was followed by a long tail. In morphology its lightcurve resembles that of a recent burst observed from GX~3+1 \citep{Chenevez2005Two-phase-X-ray}.  

\begin{deluxetable*}{llll}
\tablewidth{0pc}
\tablecaption{Bursts from Sources with Low Accretion Rates \label{tab:bursts_lowrate}}
\tablehead{\colhead{\rm Source } & \colhead{\rm Burst Duration} & \colhead{\rm Persistent luminosity} & \colhead{\rm Reference} \\
 \colhead{} & \colhead{\rm e-folding time (s)} & \colhead{\rm ($10^{36}~{\rm erg\,s^{-1}}$)} & \colhead{ }}
\startdata
\source{SAX}{1324.5}{-6313} & $6.0\pm0.1$ & $< 4\times10^{-4}$ & C02, A04 \\
\source{1RXS}{J1718.4}{-4029} & $47.5$ & $0.07$ & K00, A04, I05\\
\source{GRS}{1741.9}{-2853} & $8.8, 11.0, 16.0$ &  $0.2$ & C99, A04\\
\source{SAX}{1752.3}{-3138} & $21.9$ & $<3\times10^{-4}$ & C01, A04\\
\source{SAX}{1753.5}{-2349} & $8.9$ & $<4\times10^{-4}$ & I98, C02, A04  \\
\source{SAX}{J1806.5}{-2215} & $4.0, 9.0$ & $3$ & I98, A04\\
\source{SAX}{1818.7}{+1424} & $4.5\pm0.1$ & $<4\times10^{-4}$ & C02\\
\source{SAX}{1828.5}{-1037} & $11.2\pm0.6$ & $0.034$ & C02, A04 \\
\source{SAX}{J2224.9}{+5421} & $2.6\pm0.2$ & $<1.2\times10^{-3}$ & C02 \\
\tableline
\source{1RXS}{J1708.5}{-3219} & $\sim 300$ & $1.5$ & I05 \\
\source{SLX}{1737}{-282} & $\sim 600$ & $\leq (0.5-1.8)$  & I02\\ 
\source{SLX}{1735}{-269} & $\sim 700$ & $\sim 4$ & M05
\enddata
\tablerefs{A04: \citet{arefiev:type-i}; C99, C01: \citet{cocchi.bazzano.ea:discovery-1741,cocchi.bazzano.ea:discovery}; C02: \citet{cornelisse.verbunt.ea:bepposax}; 
K00: \citet{kaptein..ea:discovery}; I98, I02, I05: \citet{intzand98,intzand02,int-Zand2005On-the-nature-o}; M05: \citet{molkov05}.}
\end{deluxetable*}

\subsubsection{Weak Hydrogen Flashes}\label{sec:theory-bursts-low}

When the temperature in the neutron star envelope is sufficiently low, the CNO cycle becomes temperature dependent; as a result, the ignition of H becomes thermally unstable at low accretion rates \citep{fujimoto81:_shell_x}. As a first investigation of the unstable ignition of H when the atmosphere is stratified, we perform several calculations of the burst nucleosynthesis.  We approximate the cooling by a one-zone finite differencing over the envelope,
\begin{equation}
  \label{eq:one-zone}
  \CP\frac{dT}{dt} = \enuc - \ecool,
\end{equation}
where $\ecool = \rho K T y^{-2}$ and we evaluate \enuc\ from a reaction network.
Included in this network are 686 isotopes covering proton-rich nuclei up to Xe \citep[see][]{schatz.aprahamian.ea:endpoint}. The reaction rates are taken from the compilation \code{reaclib} \citep[see][and references therein]{thielemann86,rauscher00}, and consist of experimental rates and  Hauser-Feshbach calculations with the code \nonsmoker\ \citep{rauscher00}.  The initial temperature, column density, and composition are taken from the values of the quasi-static calculation at the ignition point. 
For the unstable ignition of H, the temperature in the accreted envelope is set by the flux emergent from the deeper ocean and crust.  As a result, the accretion rate at which the burst behavior changes will depend on assumptions about the heating in the crust. 

\begin{figure}[bt]
\centering
\includegraphics[width=3.4in]{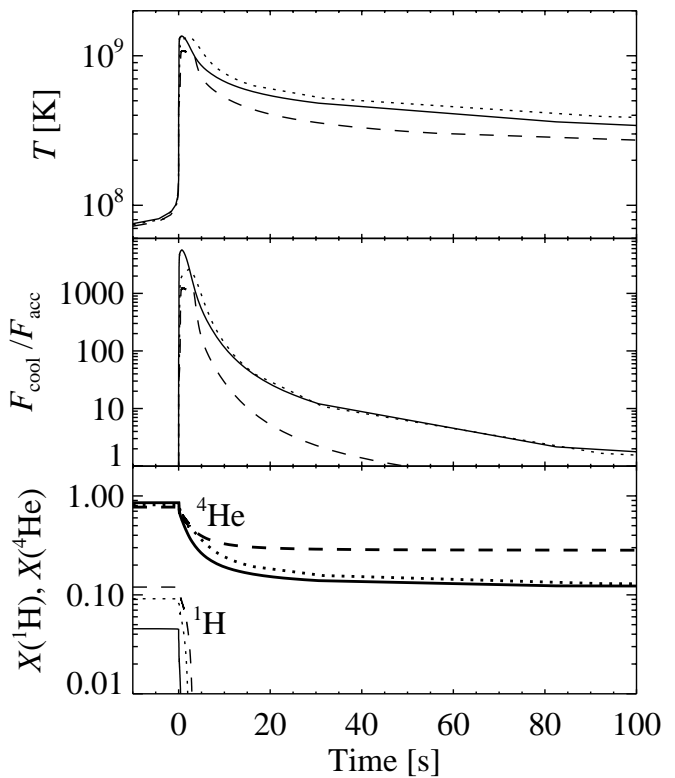}
\caption{One-zone burst calculation following unstable H ignition for three mass accretion rates: $\mdot/\medd = 9.1\ee{-4}$ (\emph{solid lines}), $1.1\ee{-3}$ (\emph{dotted lines}) and $2.3\ee{-3}$ (\emph{dashed lines}), respectively. Top panel: temperature evolution; Middle panel: the ratio of one-zone cooling flux to the accretion flux; Bottom panel: mass fraction of hydrogen (\emph{thin lines}) and $\iso{4}{He}$ (\emph{thick lines}), respectively. Diffusion and sedimentation is included. This category of H ignition triggers helium ignition and produces strong x-ray burst. }
\label{all_comp_sma.f}
\end{figure}

\begin{figure}[bt]
\centering
\includegraphics[width=3.4in]{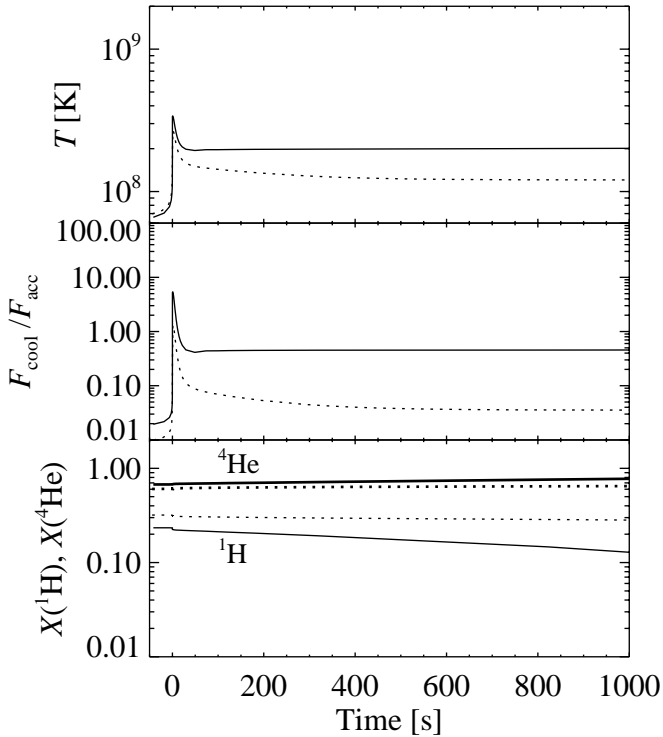}
\caption{One-zone burst calculation following unstable H ignition for two mass accretion rates: $\mdot/\medd = 5.7\ee{-3}$ (\emph{solid lines}) and $0.011$ (\emph{dotted lines}), respectively. The temperature rise is insufficient to trigger He ignition}
\label{all_comp_lma.f}
\end{figure}

We shall first describe the outcome of these one-zone calculations (eq.~[\ref{eq:one-zone}]); we will indicate below how our estimate of \ecool\ may need to be modified if sedimentation is active.  We compute the ignition conditions as described in \S~\ref{sec:evolution-ignition}, and for the first set we fix $F = 0.1\nsp\MeV\nsp\mdot/\mb$. Figure~\ref{all_comp_sma.f} shows the post-ignition evolution for accretion rates $9.1\ee{-4}\nsp\medd$ (80\nsp\gpscps; \emph{solid lines}),  $1.1\ee{-3}\nsp\medd$ (100\nsp\gpscps; \emph{dotted lines}), and $2.3\ee{-3}\nsp\medd$ (200\nsp\gpscps; \emph{dashed lines}). We plot three different quantities: the evolution of temperature (\emph{top panel}), the heat flux,
$F_{\rm cool} = y \ecool$, normalized to the accretion flux (\emph{middle panel}), and the mass fractions of hydrogen and helium (\emph{bottom panel}). We then repeat this calculation at higher mass accretion rates (Fig.~\ref{all_comp_lma.f}), $5.7\ee{-3}\nsp\medd$ (500\nsp\gpscps; \emph{solid lines}) and $1.1\ee{-2}\nsp\medd$ ($10^{3}\nsp\gpscps$; \emph{dotted lines}). 

It is immediately clear that there are two very different outcomes for the H ignition. At lower accretion rates, the rise in temperature following H ignition is sufficient to trigger a vigorous He flash with a decay timescale $\sim 10\nsp\second$ (Fig.~\ref{all_comp_sma.f} \emph{middle panel}).  At higher accretion rates, however, the flash is too weak to ignite He. As an interpretation of these results, recall that the ignition curve for the $3\alpha$ reaction has a turning point at $\yturn\approx 5\ee{7}\nsp\GramPerSc$,  which terminates the unstable branch \citep{bildsten:thermonuclear,cumming.bildsten:rotational}.  At the higher accretion rates (Fig.~\ref{all_comp_lma.f}) the ignition of H occurs at $y < \yturn$, and the local rise in temperature does not trigger unstable He ignition. In fact, the temperature rise (see Fig.~\ref{all_comp_lma.f}) is not even sufficient to initiate convection, as the radiative temperature gradient needed to carry $F_{\mathrm{cool}}$ is $d\ln T/d\ln P \approx 0.25 < (\partial\ln T/\partial\ln P)_{s}$.  In contrast, at lower accretion rates (Fig.~\ref{all_comp_sma.f}), the ignition of H occurs at $y > \yturn$, and the rise in temperature will ignite the triple-$\alpha$ reaction.  

This behavior was anticipated by \citet{fujimoto81:_shell_x}, who noted that the hydrogen-shell flash would not develop continuously into a helium-shell flash unless the accreted mass were sufficiently large. \citet{fujimoto81:_shell_x} suggested that in this case, the ignition of hydrogen would lead to temperatures needed for stable H burning and a helium flash would be triggered, as if the accretion were at higher rates. We do not find this behavior. A similar case was found in a numerical calculation \citep{ayasli82}, in which unstable H ignition raised the temperature of the envelope and gradually led, after $\approx 10^{3}\nsp\second$, to unstable He ignition. This calculation used a much smaller neutron star radius (6.57\nsp\km). As a result, the local accretion rate for unstable H ignition was much higher than ours; the reason they found an initial H flash was that they had assumed a very cold ($5.4\ee{7}\nsp\K$) crust and core. \emph{Unlike these previous works, we find a regime in which the weak H flash does not lead to sustained high envelope temperatures and a subsequent He flash.} Because our envelope cools after H ignition, the accumulated He should ignite at a much greater depth than found previously. We discuss this further in the next section.

Because sedimentation concentrates the CNO abundance into a thin layer, the expression for \ecool, which implicitly assumes that the burning is over a scale height, may underestimate the cooling rate. We find that at $\mdot = 1.1\ee{-3}\nsp\medd$ ($100\nsp\gpscps$), the CNO is concentrated into a layer of thickness $\Delta y/y \approx 0.2$. The value of \ecool\ could therefore be larger by up to $(y/\Delta y)^{2}\approx 25$. As a test, we also compute our one-zone run for this accretion rate with $y$, in the expression for \ecool, multiplied by $\xi$ so that \ecool\ is multiplied by $\xi^{2}$. We find that for $\xi \lesssim 3$, there is little change in the evolution of the burst; at $\xi = 5$, however, the peak temperature is reduced by a factor $\approx 2$ and the burst is considerably weaker. In the absence of a multi-zone calculation, we cannot determine with certainty the increase in \ecool; so long as our value is not more than a factor of $\approx 10$ too low, however, our calculations should remain valid guides to the actual behavior.

\begin{figure}[bt]
\centering
\includegraphics[width=3.4in]{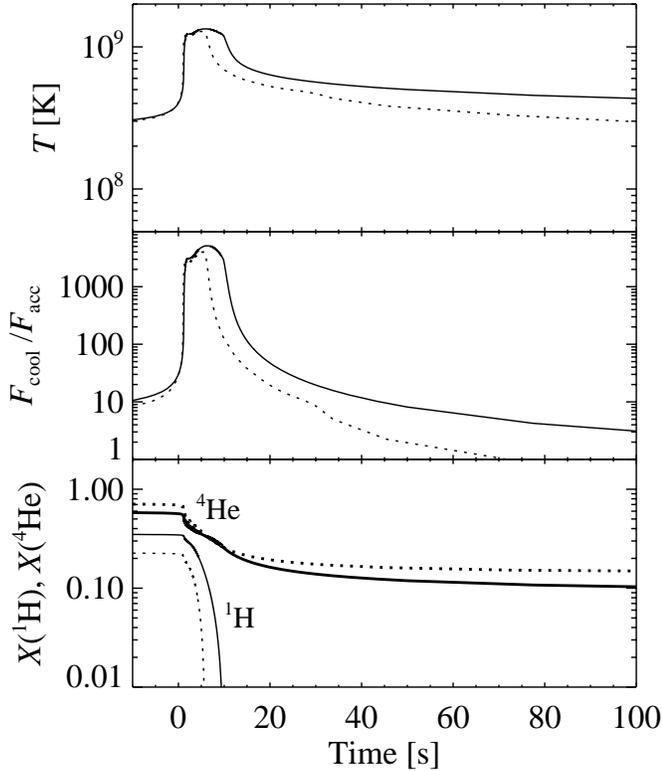}
\caption{ One-zone burst calculation following unstable H ignition for two mass accretion rates: $\mdot/\medd = 9.1\ee{-4}$ (\emph{solid lines}) and $1.1\ee{-3}$ (\emph{dotted lines}), respectively. Similar to Fig.~\ref{all_comp_sma.f}, but for cases without diffusion and sedimenation. The mass fraction of H decreases from 0.7 to 0.3 in the $10^{4}\nsp\second$ prior to He ignition.}
\label{all_comp_sma-d.f}
\end{figure}

\begin{figure}[bt]
\centering
\includegraphics[width=3.4in]{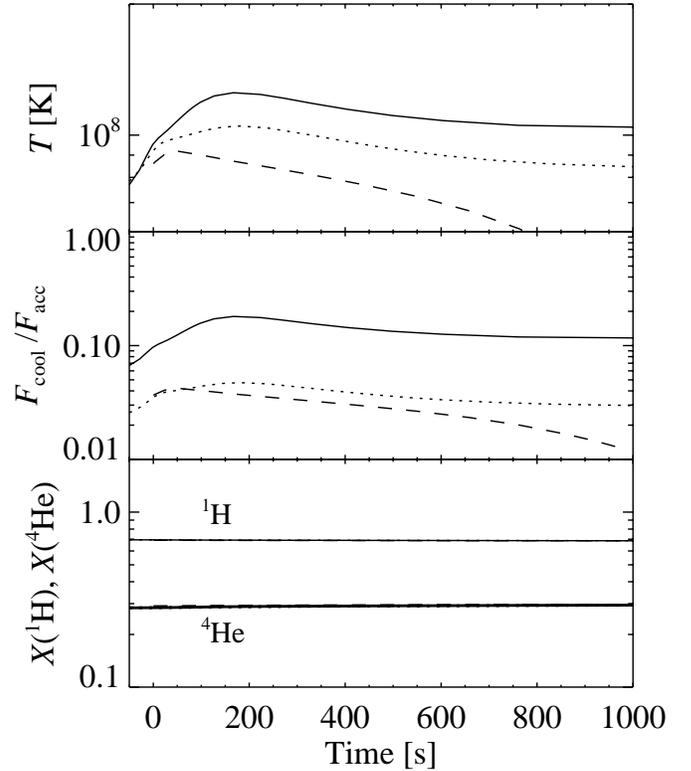}
\caption{One-zone burst calculation following unstable H ignition for three mass accretion rates: $\mdot/\medd = 2.3\ee{-3}$(\emph{solid lines}), 5.7\ee{-3} (\emph{dotted lines}) and $0.011$ (\emph{dashed lines}), respectively. Similar to Fig.~\ref{all_comp_lma.f}, but for cases without diffusion and sedimenation.}
\label{all_comp_lma-d.f}
\end{figure}

We repeated the preceding one-zone calculations for initial conditions computed without sedimentation and diffusion (Figs.~\ref{all_comp_sma-d.f} and \ref{all_comp_lma-d.f}). The strong bursts in Fig.~\ref{all_comp_sma-d.f} have a slower rise and broader peak, but decay on a similar timescale, which is expected as the ignition column is similar in both cases. Notice that the H mass fraction decreases from $\approx 0.7$  to $\approx 0.3$ in the $10^{4}\nsp\second$ before He ignition is triggered. Although the initial instability is triggered by H capture onto CNO nuclei, this quickly saturates, and the hydrogen burning rate goes to the $\beta$-limited rate $\epsilon_{\rm HCNO} = 5.8\times10^{15}Z_{\rm CNO}\nsp\ergs\usp\gram^{-1}\usp\second^{-1}$. The smaller CNO mass fraction (0.02 compared to 0.1 with sedimentation included) corresponds to a smaller energy generation rate, which requires a longer burning time and more H fuel exhaustion to reach the same He ignition temperature. The overall amount of hydrogen available for the rp-process is larger, however, which explains the broader peak. The similar decay time results from the similar amount of H left after the peak.  For the same reason (smaller energy generation rate), the H flashes that occur at higher accretion rates (Fig.~\ref{all_comp_lma-d.f}) are much weaker than those with sedimentation and diffusion included. The hydrogen depletes on a much longer timescale because of the lower abundance of CNO nuclei.  It is the abundance of CNO nuclei that sets the available heat deposition and hence the peak temperature reached during the flash, $\lesssim 10^8\nsp (Z_{\rm CNO}/0.01) \nsp\K$~\citep{ayasli82}. The difference between Fig.~\ref{all_comp_lma.f} and Fig.~\ref{all_comp_lma-d.f} is thus the amount of H consumed; sedimentation enhances the abundance of CNO at ignition and there is a sharper temperature rise.

\begin{figure}[hbt]
\centering
\includegraphics[width=3.4in]{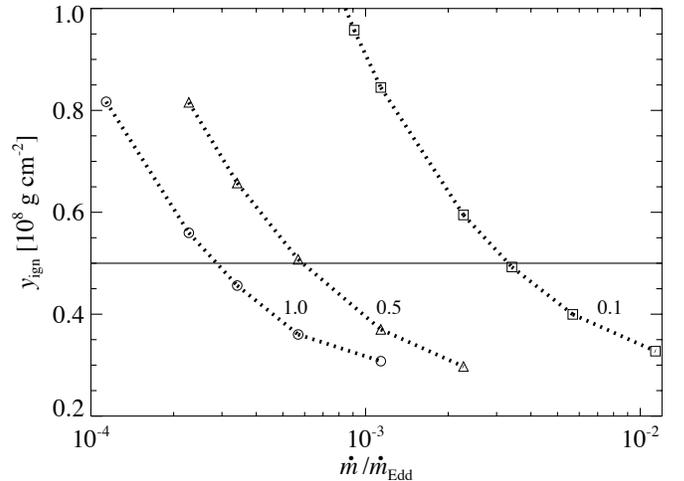}
\caption{
  Ignition column density of hydrogen burning as a function of mass accretion rates for  $F\nsp\mb/\mdot = 0.1$ (\emph{open squares}), $0.5$ (\emph{open triangles}), and $1.0\nsp\MeV$ (\emph{open circles}), respectively. The solid line indicates the minimum column for the unstable ignition of a pure He mixture \yturn. }
\label{flux_comp.f}
\end{figure}

When sedimentation is neglected, the transition accretion rate, below which the H ignition triggers vigorous bursts, decreases.  For example, in the absence of sedimentation, the H ignition at a rate $\mdot/\medd = 2.3\ee{-3}$ does not lead to ignition of the triple-$\alpha$. The reason is that the decreased He abundance,  $X(\iso{4}{He}) \sim 0.7$,  causes $\yturn$ to increase, in this case to $\yturn \approx 8\times 10^7\nsp\gram\usp\cm^{-2}$.  Under these conditions, H ignites at $y < \yturn$ for $\mdot\lesssim 10^{-3}\medd$ (cf.\ Fig.~\ref{flux_comp.f}).

We computed the ignition column for H burning for a range of $\mdot$ and $F$. Figure~\ref{flux_comp.f} displays the H ignition column density as a function of $\mdot$ for $F\mb/\mdot = 0.1$ (\emph{open squares}),  $0.5$ (\emph{open triangles}) and $1.0\nsp\MeV$ (\emph{open circles}), respectively.  We also plot a representative value of \yturn\ to guide the eye (\emph{horizontal solid line}). A larger emergent flux $F$  increases the temperature and decreases the H ignition column density. Notice that if we were to plot the H ignition column against $F$ the curves would line up; the curve $F\mb/\mdot = 1.0\nsp\MeV$ is approximately a translation 
of the curve $F\mb/\mdot = 0.1\nsp\MeV$ shifted downward by 10 in accretion rate. In the absence of steady H burning, it is the flux emergent from the crust that sets the temperature structure. 
 
\subsubsection{Accumulation of a Large He Layer}\label{sec:accumulation-He}

In these one-zone calculations of weak H flashes, the timescale for the H to be consumed is $\sim 10^{3} \nsp(0.1/Z_{\rm CNO}) (X_{\rm H}/0.3)\nsp\second < y/\mdot$.  We are unable to follow the system through a large number of repeated bursting cycles, but we note here that if subsequent H flashes do not ignite the underlying He, then a large layer of nearly pure He will accumulate.  Because our system is not burning H steadily, the temperature in the He layer is colder than if the burning were in steady-state. \emph{As a result,  a large He layer should accumulate.}

\begin{figure}[bt]
\centering
\includegraphics[width=3.4in]{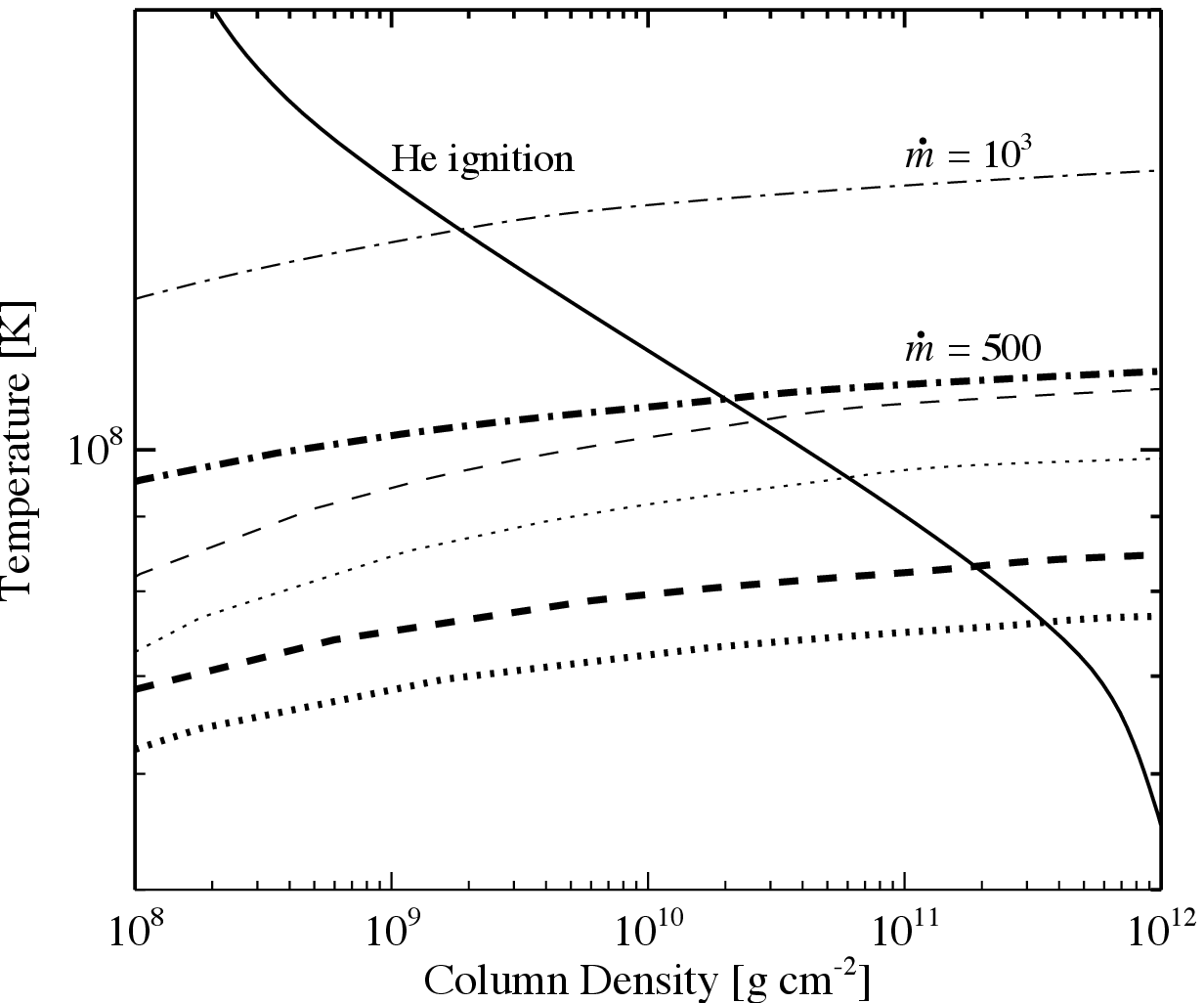}
\caption{Steady-state thermal structure of an accumulated pure helium layer for two mass accretion rates: $\mdot = 500$ ($\mdot/\medd = 5.7\ee{-3}$; thick lines) and $10^3\usp\gpscps$ ($\mdot/\medd =0.011$; thin lines), respectively. The results for three different inner fluxes, $F\nsp\mb/\mdot=0.1$ (\emph{dotted lines}), $0.2$ (\emph{dashed lines}) and  $1.0\nsp\MeV$ (\emph{dot-dashed lines}), are presented. The pure helium ignition curve (\emph{solid line}) is also shown.}
\label{evol_ign.f}
\end{figure}

To illustrate this, we show in Fig.~\ref{evol_ign.f} the thermal structure of an accumulated pure He layer for $\mdot/\medd=5.7\ee{-3}$ and 0.011. We integrated the thermal structure equations (eq.~[\ref{radiation.e}] and [\ref{heat.e}] with $\partial/\partial t\to 0$) from the column where H ignition occurred, and set the temperature there to that found in the accumulating model.  That is, we neglect heating of the H layer by the weak H flashes.  In that case, the problem becomes similar to that computed by \citet{Cumming2005Long-Type-I-X-r}.  For our purposes, we treat the flux emergent from the crust as a free parameter and show our solutions for $F \mb/\mdot = 0.1$ (\emph{dottted lines}), 0.2  (\emph{dashed lines}) and $1.0\nsp\MeV$ (\emph{dot-dashed lines}).

Superposed on this figure is the ignition curve for pure He (\emph{solid line}) defined by a one-zone stability criterion, $(\partial\epsilon_{\mathrm{nuc}}/\partial\ln T)_{P} = (\partial\epsilon_{\mathrm{cool}}/\partial\ln T)_{P}$.  The flux from the crust sets the temperature gradient in the accreted He, and we show the thermal structure for three different values,  Helium ignites at $y\approx 2\times10^{10}\textrm{--}3\times10^{11}\nsp\gram\usp\cm^{-2}$ (recurrence time $\approx 1-19\nsp\yr$) and at $y\approx2\times10^{9}\textrm{--}6\times10^{10}\nsp\gram\usp\cm^{-2}$ (recurrence time $\approx 0.06-2\nsp\yr$) for $\mdot/\medd=5.7\ee{-3}$ and $0.011$ ($500\nsp\gpscps$ and $10^3\usp\gpscps$), respectively. These results are sensitive to the flux from the deep crust, as noted by \citet{Cumming2005Long-Type-I-X-r}. The flux from the deep crust depends on the composition of the ash products and the mass accretion rates \citep{brown:nuclear,brown:superburst}. A larger inner flux increases the temperature and decreases the He ignition column. These He flashes at great depth should be similar in character to those of \source{2S}{0918}{-549} \citep{int-Zand2005On-the-possibil}. Our ignition columns for a given $F$ and $\mdot$ are less than that of \citet{int-Zand2005On-the-possibil} because our temperatures in the accreted H envelope are higher.

We were initially quite excited that energetic He bursts could also be observed from systems accreting H-rich material over a range of accretion rates.  Indeed, we note that the inferred mass accretion rates listed in the bottom part of Table~\ref{tab:bursts_lowrate} are roughly consistent with the scenario we have sketched here. There is a potential inconsistency, however.  We find weak H flashes for $\mdot/\medd \approx 10^{-2}$ for $F\nsp\mb/\mdot = 0.1\nsp\MeV$, but this low flux leads to huge He ignition depths ($y \sim 10^{11}\nsp\gram\usp\cm^{-2}$). The cooling timescale of a burst ignited at this depth is an order of magnitude longer than what is observed from the long bursts. A larger flux from the crust, which is more appropriate for low mass accretion rates \citep{brown:superburst}, increases the temperature and decreases the He ignition depth, as shown in Fig.~\ref{evol_ign.f}. This increased flux decreases, however, the range of mass accretion rates for which weak H flashes can occur. For example, at $F  \mb/\mdot = 1.0\nsp\MeV$, the range of mass accretion rates for weak H flashes decreases to $3 \times 10^{-4}\usp \medd \lesssim \mdot \lesssim 10^{-3}\usp \medd$ (see Fig.~\ref{flux_comp.f}), which is lower than the accretion rate inferred for these bursters. Given the difficulty in inferring mass accretion rates from observed fluxes and the crudeness of our one-zone calculations, it is uncertain whether this mismatch is a serious problem for our model or not.

\subsection{\iso{12}{C} Production and Implications for Superbursts}\label{sec:carbon-production}

Superbursts have been detected from 10 sources to date \citep{cornelisse.ea:longest,.kuulkers.ea:bepposax,kuulkers:gx31_super,strohmayer.brown:remarkable,wijnands:recurrent,.cornelisse.ea:superbursts,Kuulkers2005Probable-superb}. The mean recurrence time of these superbursts is $\approx 1.5\nsp\yr$ \citep{.cornelisse.ea:new}, and
the majority of superbursters discovered to date accrete hydrogen-rich matter at rates 0.1--0.3\nsp\medd.
\citet{cumming.bildsten:carbon} proposed that the unstable ignition of a small amount ($\sim 10\%$ by mass) of \iso{12}{C} at densities $\lesssim 10^{9}\nsp\grampercc$ would suffice to power these bursts.  This scenario predicts recurrence times, energetics, and cooling timescales \citep{cumming.macbeth:thermal} that are in good agreement with observations provided that the deep crust of the neutron star is sufficiently hot \citep{brown:superburst,cooper.narayan:theoretical,Cumming2005Long-Type-I-X-r}. A challenge to this scenario is the lack of \iso{12}{C} produced in both one-zone and multi-zone calculations of X-ray bursts with full reaction networks \citep{schatz.aprahamian.ea:endpoint,woosley.heger.ea:models,Fisker2005Extracting-the-,Fisker2005The-reactions-a}, although a greater abundance of carbon may be produced if the burning at lower accretion rates were stable \citep{Cumming2005Superbursts:-A-,Cooper2005On-the-Producti}.

As shown by Fig.~\ref{abund_mdot.f}, sedimentation decreases the H abundance and increases the CNO abundace at the base of the accreted envelope, \emph{even at accretion rates $\gtrsim 0.1\nsp\medd$}. It is the ratio of protons to heavier ``seed'' nuclei that determines the run of the rp-process \citep[for a succinct discussion, see][]{schatz99}.  \emph{In the absence of mixing, sedimentation lowers the proton-to-seed ratio and decreases the mean mass of the rp-process ashes.} As we noted in \S~\ref{sec:bursts-at-low}, during the burst rise a convective region is established which moves outward in mass and mixes the envelope; this prevents us from giving a definitive statement about how the stratification of the envelope changes the burst ashes. Just to motivate the problem, we perform a one-zone calculation (\S~\ref{sec:bursts-at-low}, eq.~[\ref{eq:one-zone}]) of the burst nucleosynthesis. The initial temperature, column density, and composition are taken from the values of the quasi-static calculation at the ignition point for a mass accretion rate of $\mdot=0.11\nsp\medd$. We checked our calculation against the results of \citet{schatz.aprahamian.ea:endpoint}; for similar conditions, we find a broad distribution of isotopes with $A = 60\textrm{--}100$, but our distribution peaks at $A\approx 64$.

\begin{figure}[hbt]
\centering
\includegraphics[width=3.4in]{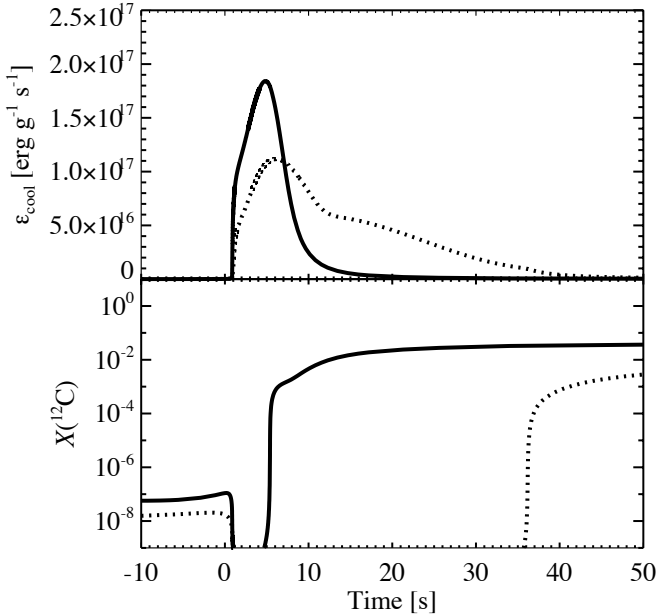}
\caption{The one-zone cooling rates \ecool\ (eq.~[\ref{eq:one-zone}]) and the mass fractions of  \iso{12}{C} for a one-zone burst calculation at $\mdot = 0.11\medd$. The initial composition is taken from the bottom of the fuel layer at the point \iso{4}{He} ignites, for the calculation with (\emph{solid}) and without (\emph{dotted}) sedimentation and diffusion.}
\label{burst_1e4.f}
\end{figure}

\begin{figure}[hbt]
\centering
\includegraphics[width=3.4in]{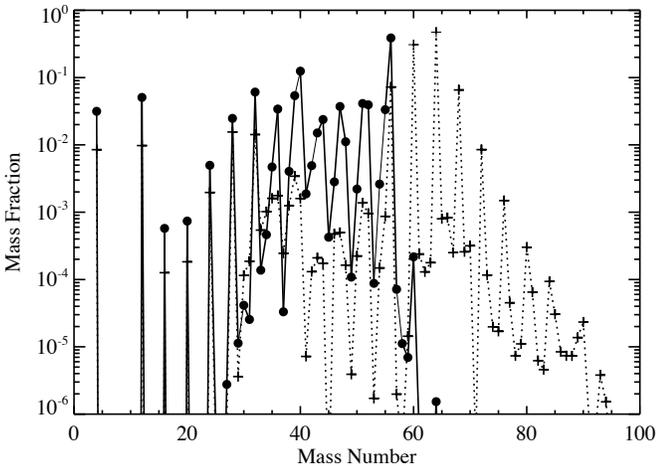}
\caption{The ashes produced by the one-zone burst calculation following unstable He igntion for $\mdot = 0.11\medd$. The initial composition is taken from the bottom of the fuel layer at the point \iso{4}{He} ignites, for the calculation with (\emph{dots, solid lines}) and without (\emph{crosses, dotted lines}) sedimentation and diffusion.}
\label{abund_end_1e4.f}
\end{figure}

Figs.~\ref{burst_1e4.f} and \ref{abund_end_1e4.f} show the results of this one-zone calculation, for ignition conditions calculated both with (\emph{solid lines}) and without (\emph{dotted lines}) a depressed hydrogen abundance. In Fig.~\ref{burst_1e4.f} we plot \ecool\ (a crude measure of the burst lightcurve; \emph{top panel}) and the mass fraction of \iso{12}{C}, while Fig.~\ref{abund_end_1e4.f} shows the composition of the ashes. The temperature rises rapidly for $t<5\nsp\second$ during the helium flash. Helium burns via triple-alpha reaction and produces carbon, which is consumed via $\iso{12}{C}(p,\gamma)\iso{13}{N}$. As noted by \citet{Schatz2003Nuclear-physics}, \iso{12}{C} is only abundant after the H has been consumed. When $X(H)$ is decreased by a factor $\approx 2$ (Fig.~\ref{abund_mdot.f}), \iso{12}{C} is produced earlier and in greater abundance. In addition, the lower proton-to-seed ratio shortens the burst and shifts the composition to lower mass numbers. For this accretion rate, $X(\iso{12}{C})$ is increased by a factor of 5 (0.01 to 0.05) in the ashes. We repeated this calculation for $\mdot=0.057\nsp\medd$ and $\mdot=0.23\nsp\medd$, and found that $X(\iso{12}{C})$ increased by a factor $\approx 3$ and $\approx 7$, respectively: the overproduction of \iso{12}{C} is greater at higher accretion rates. 

As a check, we also performed the run at $\mdot=0.11\nsp\medd$ with varying resistance coefficients $\K_{ij}$. As discussed in Appendix~\ref{resistance}, the $K_{ij}$ are very uncertain (see Fig.~\ref{resist_H_He.f}). Reducing $K_{ij}$ by a factor of 5 enhances the sedimentation of He and CNO elements and leads to an increase in the \iso{12}{C} mass fraction after the He flash by a factor of $\sim 2$ ($X(\iso{12}{C}) = 0.09$). Conversely, increasing $K_{ij}$ by a factor of 5 (smaller $D_{ij}$), decreases $X(\iso{12}{C})$ by a factor of 2 ($X(\iso{12}{C})=0.03$).

\section{Summary and Conclusions}\label{sec:disc-concl}

Using a simplified numerical model of an accreting neutron star envelope that allows for differential isotopic velocities, we have undertaken a first study of the effects of sedimentation and diffusion on the unstable ignition of hydrogen and helium on the surfaces of accreting neutron stars, and have investigated the outcome of unstable hydrogen ignition using simple one-zone models. Our principal conclusions are:
\begin{enumerate}
\item The effect of sedimentation changes the conditions at H/He ignition even for $\mdot \gtrsim 0.1\medd$.

\item There is a range of accretion rates for which unstable H ignition does not trigger a He flash.  This range depends on the flux from the deep crustal heating and the degree of settling of He and CNO nuclei. For $F\nsp(\mb/\mdot) = 0.1 (1.0) \nsp\MeV$, the range is $3\times 10^{-3}\nsp\medd\lesssim\mdot\lesssim 10^{-2}\nsp\medd$ ($3\ee{-4}\nsp\medd\lesssim\mdot\lesssim 10^{-3}\nsp\medd$). In contrast to previous calculations \citep{fujimoto81:_shell_x,ayasli82}, we find that the flash does not lead to quasi-steady H burning  that develops after $\approx 1000\nsp\second$ into a He flash. We speculate that successive weak H flashes can lead to the accumulation of a large reservoir of He; this may explain the long bursts observed from some sources. For accretion rates lower than this range, H ignition leads to a strong mixed H/He flash. The rise time of the flash and the burst duration, however, depend on the abundance of H and CNO nuclei at the ignition.  The sense of this (a transition from vigorous bursting to weak H flashes interrupted by less frequent, strong He flashes) agrees with the observations of bursts at low $\mdot$, although our estimates of the accretion rate at which this transition happens does not match the observed value.

\item For $\mdot\gtrsim 0.1\nsp\medd$, the partial stratification of the envelope means that the base of the accreted layer is deficient in hydrogen, relative to the case where complete mixing is assumed. Because a convective zone is established during the onset of the burst, we cannot say for certain what effect, if any, this will have on the outcome of the burst. We do note, however, that in the absence of mixing, the production of \iso{12}{C} is enhanced, by a factor of 5.
\end{enumerate}

There are several items that are open for further investigation. First is the interplay between convection and the rp-process breakout. At these accretion rates, the unstable burning of \iso{4}{He} and the breakout from the HCNO cycle drives a convective zone outwards that encompasses $>90\%$ of the accreted envelope \citep{Weinberg2005Exposing-the-Nu}. This would rapidly mix the envelope and ``reset'' the amount of protons available for capture onto the endpoint of the $(\alpha,p)(p,\gamma)$. Observationally, the onset of stability is closer to $\mdot \approx 0.3\nsp\medd$ and not $\medd$ as found by most one-dimensional calculations. In superbursting sources, much of the accreted fuel must be consumed stably to account for the low values of $\alpha \equiv \textrm{(persistent fluence)}/\textrm{(burst fluence)}$ \citep{.cornelisse.ea:new}. Close to stability, the growth of the convection zone is reduced \citep*[see][for a discussion of the behavior near stability; note that in their calculation this occurs at close to the Eddington limit]{Heger2005Millihertz-quas}. The extent to which sedimentation affects the nucleosynthesis in this regime therefore remains an open question. A second task is to perform one-dimensional calculations of unstable H ignition under conditions for which the ignition does not trigger a He flash. This is critical for determining how sedimentation modifies the finite-difference approximation for the cooling rate, as described in \S~\ref{sec:theory-bursts-low}. Following successive weak flashes is necessary to determine how large a mass of He can in fact be accumulated. Finally, as discussed in \S~\ref{sec:carbon-production}, our results depend on the values of resistance coefficients. A further investigation of resistance coefficients 
at $\Gamma \gtrsim 1$ will help refining our results. 

\acknowledgements

It is a pleasure to thank Phil Arras, Lars Bildsten, Randy Cooper, Andrew Cumming, Philip Chang, Ami Glasner, Alexander Heger, Jean in 't Zand, and Hendrik Schatz for stimulating discussions and helpful comments. We also thank the referee for a constructive critique. This work is supported in part by the \textbf{J}oint \textbf{I}nstitute for \textbf{N}uclear \textbf{A}strophysics under NSF-PFC grant PHY~02-16783, and by the Department of Energy under grant B523820 to the Center for Astrophysical Thermonuclear Flashes at the University of Chicago. EFB is supported by the NSF under grant AST-0507456; JWT acknowledges support from the U.S. DOE, under contract No. W-31-109-ENG-38. 

\appendix

\section{Resistance Coefficients}\label{resistance}

In this appendix, we discuss our choice of the resistance coefficients $K_{ij}$ for a multicomponent plasma. When the plasma is sufficiently rarefied, particle pairs interact via a screened Coulomb potential, and the resistance coefficient in this limit is
\begin{equation}
  K_{ij}^0 = \frac{2^{3/2}}{3}\pi^{1/2}\mu_{ij}^{1/2} \left(\frac{\mb}{\kB T}\right)^{1/2}
  \frac{e^4Z_i^2Z_j^2}{\kB T} n_i n_j \ln (1 + x_{ij}^2).
\label{kij0.e}
\end{equation}
Here $\mu_{ij} = A_iA_j/(A_i + A_j)$ and $x_{ij} = 4\lD\kB T/(e^{2}Z_i Z_j)$, with $\lD = \left[\kB T/(4\pi e^{2}\sum_i Z_i^2 n_i)\right]^{1/2}$ being the Debye length. The sum in the expression of \lD\ is over all species in the plasma, including electrons. Using
\begin{equation}
\label{e.plasG}
\Gamma =  \frac{\Zbar[2] e^{2}}{a \kB T} 
\end{equation}
and the ion plasma frequency
\begin{equation}
\label{e.plas-freq}
\wpl^{2} = \left(\frac{\Zbar[2] e^{2} n}{\Abar m_\mathrm{u}}\right) ,
\end{equation}
we rewrite equation~(\ref{kij0.e}) in terms of $\Gamma$ and \wpl,
\begin{equation}
\label{e.kij0-alt}
\K_{ij}^{0} = \left(6\pi\right)^{-1/2}
\left(\frac{\Abar^{2}}{A_{i}A_{j}}\right)
\left(\frac{Z_{i}Z_{j}}{\Zbar[2]}\right)^{2} \left(\frac{\mu_{ij}}{\Abar}\right)^{1/2} 
\wpl\rho \Gamma^{3/2} \Lambda.
\end{equation}
Here $\Lambda = \ln(1+x_{ij}^{2})$. This form of $K_{ij}^{0}$ is for our definition for $\Gamma$ (eq.~[\ref{e.plasG}]) .

When the plasma becomes strongly coupled, $\lD < a$ or equivalently, when $\Gamma \gtrsim 1$, the resistance coefficient can no longer be defined in terms of a sequence of binary collisions. This regime is important for studies of white dwarf cooling \citep[see][]{deloye.bildsten:gravitational}, and there have been numerous attempts to calculate the binary diffusion coefficient $D_{ij}$ from molecular dynamics simulations \citep[for a recent example, see][]{Daligault2005Semiclassical-m}. The resistance coefficient is related to $D_{ij}$ by $D_{ij} = (k_{\rm B} T/K_{ij})(n_i n_j/n)$, where $n = \sum_s n_s$. \citet{hansen.joly.ea:self-diffusion} found that
\begin{equation}\label{e.HJP}
K_{ij} = 0.11 \rho\wpl\Gamma^{0.34}.
\end{equation}
\citet{bildsten.hall:diffusion} used the Stokes-Einstein relation with the shear viscosity estimated from  molecular dynamics simulations \citep{donko.nyiri:viscosity} to obtain a similar relation,
\begin{equation}\label{e.BH}
K_{ij} = 0.15 \rho\wpl\Gamma^{0.3}.
\end{equation}
Both of these fits are applicable in the regime $\Gamma > 1$. For a multi-species plasma, there are no well-defined binary diffusion coefficients in the regime of $\Gamma > 1$. \citet{hansen.joly.ea:self-diffusion} noted that a linear combination of the self-diffusion coefficients computed for a OCP gave a first-order accurate estimation of the binary diffusion coefficient $D_{ij}$; when written in terms of the resistance coefficients, this becomes
\begin{equation}\label{eq:K-mix}
K_{ij} = \frac{K_{ii} K_{jj}}{K_{ii} + K_{jj}}.
\end{equation}
In applying this formula, we also use $a_{\mathrm{p}}$ in place of $a$, where $4\pi a_{\mathrm{p}}^{3}/3 = \Zbar/n_{e}$ in computing the $K_{ii}$. This follows from the Stokes-Einstein relation \citep{bildsten.hall:diffusion}.

The long-range cutoff for the Coulomb potential in equation~(\ref{e.kij0-alt}) does not hold when $a > \lD$. \citet{muchmore:diffusion} proposed redefining $x_{ij}$ to use $\max(\lD,a)$; using this definition and equation~(\ref{e.plasG}), we have
\begin{equation}\label{eq:xij}
  x_{ij} = \cases{%
      4\frac{\langle Z^{2}\rangle}{Z_i Z_j}
      \frac{1}{\Gamma}\left[\frac{1}{3\Gamma\left(1+\Zbar/\Zbar[2]\right)}\right]^{1/2}, & $\lD > a$ \cr
      4\frac{\langle Z^{2}\rangle}{Z_i Z_j} \frac{1}{\Gamma}, & 
     $\lD \le a$} .
\end{equation}
The two expressions match at $\Gamma = \onethird(1+\Zbar/\Zbar[2])^{-1}$. The factor $1+\Zbar/\Zbar[2]$ accounts for $\lD$ including the electrons in the sum over charged species. \citet{fontaine.michaud:diffusion} numerically computed binary diffusion coefficients using a Thomas-Fermi potential at high densities $x_{ij} \lesssim 1$ and a Debye-H\"uckel potential at low densities. A convenient fit to their results is \citep{iben.macdonald:effects} 
\begin{equation}
  K_{ij} = 1.6249 \frac{\ln\left[1 
      + 0.18769 x_{ij}^{1.2}\right]}{\ln\left(1 + x_{ij}^2\right)} K_{ij}^0,
\label{kij1.e}
\end{equation}    
where $x_{ij}$ is computed from equation~(\ref{eq:xij}).

\begin{figure}[t]
\centering
\includegraphics[width=4.0in]{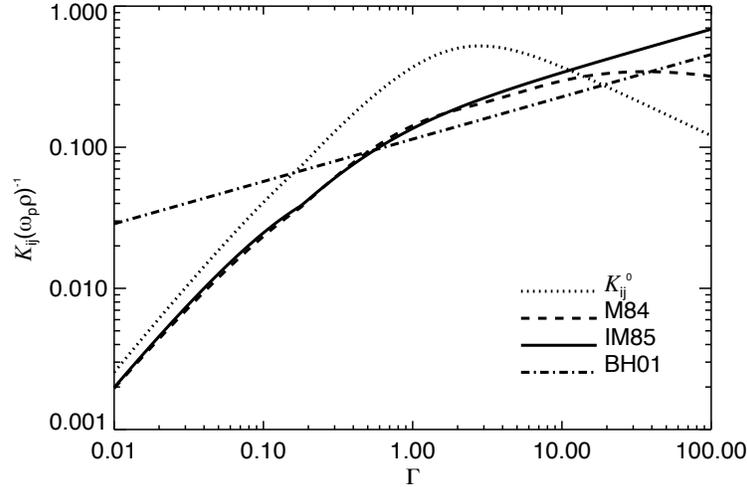}
\caption{\label{resist_H_He.f}
Normalized resistance coefficients for a H/He mixture with $X(\iso{1}{H})=0.7$ and $X(\iso{4}{He})=0.3$. We show the result for the low-density limit $K_{ij}^{0}$ (\emph{dotted line}) as well as the calculation of \citet{muchmore:diffusion} (\emph{dashed line}; M84), the fit of \citet{iben.macdonald:effects} (\emph{solid line}; IM85), and the Stokes-Einstein relation with the viscosity inferred from OCP molecular dynamics simulations \citep[\emph{dot-dashed line}; BH01]{bildsten.hall:diffusion}. }
\end{figure}

Figure~\ref{resist_H_He.f} shows the normalized resistance coefficients $K_{ij}(\wpl\rho)^{-1}$ as a function of $\Gamma$ for a two-isotope plasma with $X(\iso{}{H})=0.7,\;X(\iso{4}{He})=0.3$. We show $K_{ij}$ computed from a Debye-like potential with a screening length $\max(\lD,a)$ \citep[\emph{dashed line}]{muchmore:diffusion}; from a Thomas-Fermi potential at higher densities \citep[\emph{solid line}]{iben.macdonald:effects}; and from the Stokes-Einstein relation with the viscosity inferred from molecular dynamics simulations \citep[\emph{dot-dashed line}]{bildsten.hall:diffusion}. We also show $K_{ij}^{0}$ (\emph{dotted line}) for comparison. In the regime $0.1\lesssim \Gamma \lesssim 10$, the fit of \citet{iben.macdonald:effects} agrees well with the calculation of \citet{muchmore:diffusion}, and has the correct scaling with $\Gamma$ in the low- and high-density limits, where it differs by $\lesssim 40\%$ from the fit of \citet{bildsten.hall:diffusion}. It also agrees well with the calculation of \citet{paquette.pelletier.ea:diffusion} over the range $0.1 < \Gamma < 10.0$. For the conditions of interest in this paper, generally $\Gamma\lesssim 1$, we therefore adopt this fit throughout this paper, with the caveat that the resistance coefficients are generally uncertain in the range $\Gamma \gtrsim 1$.

\bibliographystyle{apj}
\bibliography{ms}

\end{document}